\documentclass[10pt]{article}

\usepackage{compozition}
\usepackage{bm}
\usepackage{graphicx}
\usepackage[all]{xy}
\usepackage{amsmath}

\title{\textbf{The Construction Interpretation: Conceptual Roads to Quantum Gravity}} 
\author{Lucien Hardy\\
\textit{Perimeter Institute,}\\
\textit{31 Caroline Street North,}\\
\textit{Waterloo, Ontario N2L 2Y5, Canada}}
\date{}

\begin{document}

\maketitle

\begin{abstract}
In the first part of this paper I propose the Construction Interpretation of the Quantum. The main point of this interpretation is that (unlike previous interpretations) it is not actually an interpretation but rather a methodology aimed to encourage a conceptually driven construction of a theory of Quantum Gravity.  In so doing we will, I hope, resolve the ontological problems that come along with Quantum Theory.  In the second part of this paper I offer a particular perspective on the theory of General Relativity and set out a path of seven steps and three elements corresponding (more-or-less) to the path that Einstein took to this theory.   In the third part I review a particular operational approach to Quantum Field Theory I have been developing - the operator tensor formulation.  The physicality condition (this ensures that probabilities are between 0 and 1 and that choices in the future cannot effect the past) is discussed in some detail in this part and also in an appendix.  In the fourth part of the paper I set out one particular conceptually driven possible road to Quantum Gravity. This approach works by establishing seven steps and three elements for a theory of Quantum Gravity that are analogous to those of General Relativity.  The holy grail in this approach is to generalize the physicality conditions from Quantum Field theory (which presume definite causal structure) to the new situation we find ourselves in in Quantum Gravity (where causal structure will be indefinite). Such conditions, in the present approach, would be analogous to the Einstein Field Equations. In the fifth part of the paper I propose the quantum equivalence principle whereby it is always possible to transform to a quantum reference frame such that we have definite causal structure in the vicinity of any given point. This equivalence principle suggests another possible road (albeit more speculative) to Quantum Gravity in even closer analogy to the path Einstein took to General Relativity.
\end{abstract}

\pagebreak

\tableofcontents

\pagebreak

\section{Prelude}

In 1687 Newton published his  Philosophi\ae \kern 0.125em 
Naturalis Principia Mathematica \cite{newton1999principia} in which he wrote down his law of universal gravitation
\[ F = G\frac{M_1M_2}{r^2}  \]
The force between two masses depends on the instantaneous distance, $r$, between them.  This was regarded as deeply unsatisfactory by Newton and others since it constitutes action at a distance without any mediating substance. There were numerous attempts to explain this gravitational attraction in mechanical terms.  Netwon himself (in letters to Oldenburg and Boyle \cite{burtt1925metaphysical}) considered an aether model in which the attractive force was due to the decrease of pressure when fluid passes at speed (later the effect was formalized by Bernoulli).  Descartes had a model \cite{sep-descartes-physics} in which there are vortices in fine matter (basically an aether) which catch rough matter (such as the planets) up in their motion pushing this rough matter towards the center of the vortex.  Another model (originally due to Nicolas Fatio de Duillier \cite{zehe1983gravitationstheorie} and reinvented by George-Luis Le Sage (see Poincar{\'e}'s exposition \cite{poincare1914science}) among others) has a flux of particles with equal density in all directions except where some massive object causes a shadow. Any other massive object in this shadow will feel unequal pressure and experience a net force towards the first mass.

How did these issues with Netwon's law of universal gravitation actually get resolved?  The answer came 228 years \cite{einstein1916grundlage} later when Einstein successfully solved a problem (let us call it the problem of Relativistic Gravity) to find a theory that limits to Newtonian Gravity on the one hand and to special relativistic field theories (such as Maxwell's electromagnetism) on the other
\[
\text{Newton Gravity} \longleftarrow \text{Relativistic Gravity} \longrightarrow \begin{array}{l} \text{Special Relativistic} \\  ~~~\text{Field Theories} \end{array}
\]
He called the theory of relativistic gravity he found \emph{General Relativity} because it involved consideration of transformations between general coordinate systems. For speeds small compared with $c$,  General Relativity limits to Newtonian Gravity.  For small masses it limits to the case of Special Relativistic Field Theory.  It is important to note, however, that General Relativity fits in with neither the framework of Newtonian Gravity nor of Special Relativistic Field Theories.  There are corrections to Newtonian Gravity that kick in for speeds close to that of light and to Special Relativistic Field Theories that kick in when gravitationally induced curvature is big.  There is a certain even-handedness: General Relativity modifies both Newtonian Gravity and Special Relativistic Field Theories.

In the light of General Relativity we might ask what the correct interpretation of Newtonian Gravity is.  If we take the $c\rightarrow\infty$ limit of General Relativity the actual model we get (which agrees with the predictions of Newton's gravitational theory) is something called the Newton-Cartan formalism \cite{cartan1922equations}.  In this formalism space and time separate out each having their own metrics.  These metrics are the \lq\lq stuff" that communicate the gravitational force between distant masses. The force is still instantaneous (which is fine as we have $c\rightarrow\infty$) but now it can be regarded as a contact force (curvature in the metrics mediate the force).

The Newton-Cartan formalism is of academic interest but it is not the correct description of the world.  It is, incidentally, difficult to imagine arriving at the Newton-Cartan picture without first having General Relativity.  The deeper concept is curvature of spacetime - a radically new idea from the perspective of Newtonian physics.  Einstein did not go through the intermediary of the Newton-Cartan formalism or any of the mechanical models of gravitation mentioned above to get to General Relativity. Rather, he followed a conceptual approach (which we will discuss in some detail later).

Many of the attempts to provide interpretations of Quantum Theory look to me like the mechanical models of gravitation.  They are, at best, an attempt to guess the analogue of the Newton-Cartan formalism without first attempting to solve the deeper problem of Quantum Gravity.  The problem of Quantum Gravity can be represented as follows.
\[
\text{General Relativity} \longleftarrow \text{Quantum Gravity} \longrightarrow \begin{array}{l} \text{Special Relativistic} \\  ~~~\text{Quantum Field Theories} \end{array}
\]
The problem is to find a theory that limits to General Relativity on the one hand and Quantum Theory (in particular, Quantum Field Theory) on the other hand. This clearly has a similar shape to the problem of Relativistic Gravity.  We should, then, be prepared to take an evenhanded approach to Quantum Gravity wherein both General Relativity and Quantum Field Theory are modified.

While there were many twists and turns along the way, Einstein eventually found the right path to understanding Gravity.  This involved solving the problem of Relativistic Gravity.  The purpose of this paper is to nudge researchers working on the foundations of Quantum Theory along a similar kind of path but, this time, guided by attempting to solve the problem of Quantum Gravity employing the conceptual tools developed in foundations.


\part{The Construction Interpretation}

\section{More than merely academic}

In view of the preceding remarks, I wish to propose \emph{The Construction Interpretation} of the Quantum.  The Construction Interpretation is distinct from other interpretations in that it is not actually an interpretation at all. Rather, it is a \emph{methodology} aimed at obtaining an interpretation through dedicating ourselves to a conceptually driven construction (hence the name) of a theory of Quantum Gravity.

The great danger facing existing interpretations is that they will overturned by progress in physics.  In particular, we expect Quantum Theory (and, indeed, General Relativity) will eventually be replaced by a new theory - a theory of Quantum Gravity.  It is possible, likely even, that this theory will be so different from Quantum Theory as to make existing interpretations redundant. They will still be of academic interest to historians of science but they will not be taken seriously by practising physicists as providing an actual description of the world.  If Quantum Gravity requires a radical departure from existing theories, the most interpretations of Quantum Theory can aspire to is to be the correct limiting version of Quantum Gravity just as the Newton-Cartan formalism is the correct limit of General Relativity.

We could be driven to despair.  If future developments in physics are likely to completely overturn all our hard work then what is the point in pursuing a conceptual understanding of Quantum Theory at all? Few physicists wish to dedicate themselves to merely providing amusement to future historians.  The point of The Construction Interpretation is to take the noble instincts that lead us to attempt to understand Quantum Theory in conceptual terms and re-purpose them to the problem of \emph{constructing} a theory of Quantum Gravity.  The hope (though, admittedly, this could turn out to be in vain) is that the correct theory of Quantum Gravity will (like General Relativity) suggest its own ontology and so we will, naturally, arrive at an interpretation.  Thinking about foundational issues in the light of the concrete problem of constructing a theory of Quantum Gravity will, I hope, change our perspective in a useful and productive way.  If we are genuinely interested in obtaining the best understanding of Quantum Theory then this strategy is, I suggest, more likely to lead to progress than presenting interpretations of Quantum Theory as complete and finished works.

\section{The tenets}

No respectable interpretation of the Quantum (even a non-interpretation like the present one) is without basic tenets. So, without further ado, here are the five tenets of \emph{The Construction Interpretation}.
\begin{description}
\item[Driven.]  The Construction Interpretation is driven by the intent to construct a theory of Quantum Gravity. As such, it is opportunistic in that it will adopt and abandon ideas (and, possibly, readopt them) as serves this goal.
\item[Conceptual.]  The Construction Interpretation re-purposes concepts and principles taken from the foundations of Quantum Theory and the foundations of General Relativity for the primary purpose of theory construction.
\item[Operational and ontological.]  Operational methodologies relating to actions and observations of agents are valid as are ontological methodologies which attempt to make assertions about what the underlying reality is.
\item[Frameworks and principles.]  The Construction Interpretation uses mathematical frameworks based on operational and/or ontological ideas as a place to impose different concepts and principles to get traction on the construction of a new physical theory.
\item[Non-interpretation.] The Construction Interpretation is not actually an interpretation, but rather, a methodology aimed at finding a theory of Quantum Gravity. As such, the Construction Interpretation is pluralistic - it encourages many different conceptually driven approaches.
\end{description}
My hope is that a theory of Quantum Gravity, once so constructed, will suggest its own interpretation (just as General Relativity does as discussed in Sec.\ \ref{sec:GRsuggests}).

In Parts \ref{part:GR} and \ref{part:QFT} I will present my own takes on  General Relativity and Quantum Field Theory respectively.  This is done in preparation for Part \ref{part:roadtoQG} and Part \ref{part:thequantumep} which present possible roads to a theory of Quantum Gravity.  In accord with the pluralistic exhortations of the last tenet above, I want to emphasize that these are only two possibilities and other conceptually driven approaches are encouraged.


\part{General Relativity}\label{part:GR}

\section{The problem of Relativistic Gravity}

In the early part of the last century physicists faced the problem of Relativistic Gravity.  As discussed in the prelude, this is to find a theory that gives rise to Newtonian Gravity on the one hand and Special Relativistic Field Theory (such as described by Maxwell's equations) on the other under appropriate limits.
\[
\text{Newton Gravity} \longleftarrow \text{Relativistic Gravity} \longrightarrow \begin{array}{l} \text{Special Relativistic} \\  ~~~\text{Field Theories} \end{array}
\]
This problem was solved in the years leading up to 1915 by Albert Einstein in the form of the theory of General Relativity.  Unlike Quantum Theory, there is a clear and widely agreed upon interpretation of General Relativity (which we will discuss in Sec. \ref{sec:elementsofGR}).  General Relativity represents a radical new take on reality (in some ways more radical than that of Quantum Theory).   Here I will clarify what, exactly, the theory of General Relativity along with its interpretation is.  Also, I will provide a simplified sketch of the conceptual steps that Einstein took to arrive at the theory.

\section{Einstein's path to General Relativity}\label{sec:EinsteinPath}

In setting up the framework for General Relativity Einstein followed a particular path which I will schematize in $7$ steps $+$ $3$ elements.  The seven steps take us to a framework for physical theories (that of tensor fields defined on a manifold satisfying local field equations).
\[
\begin{array}{ll}
{}          & \fbox{\text{Equivalence principle} }  \\
\rightarrow & \fbox{\text{No global inertial reference frame}} \\
\rightarrow & \fbox{\text{General coordinates}}  \\
\rightarrow & \fbox{ \text{Local physics}}   \\
\rightarrow &\fbox{\text{Laws expressed by field equations}}\\
\rightarrow &\fbox{\text{Local tensor fields based on tangent space}}\\

\rightarrow & \fbox{\text{Principle of general covariance}}
\end{array}
\]
The three elements (discussed in more detail in Sec.\ \ref{sec:elementsofGR}) single out General Relativity (one particular theory that might be formulated in this framework).
\[
\begin{array}{ll}
\rightarrow & \fbox{\text{Prescription for turning SR field equations into GR field equations} }\\
\rightarrow & \fbox{\text{Addendum: The Einstein field equations} } \\
\rightarrow  &\fbox{\text{An interpretation} }
\end{array}
\]
We will see that there is an analogous path to a possible framework for Quantum Gravity. Analogy is a potentially powerful source of inspiration in theory construction.  To be in a good position to understand the analogous steps I propose, I will outline these steps and elements in the next two Sections.  It is worth saying that the above path is intended to be schematic.  It is an oversimplification of all the twists and turns in the actual historical path that Einstein took.

\section{Seven steps to a framework}\label{sec:sevensteps}

The equivalence principle states that the laws of physics are the same locally in any freely falling frame of reference. Einstein saw that the equivalence principle implies that we cannot set up a global inertial reference frame as in Special Relativity.  This forces the use of general coordinates.  These general coordinates can be thought of as charting some part of a manifold, $\mathcal{M}$.  To cover all of the manifold we may need multiple such charts.  These are the first three steps.

Once we have a coordinate system we introduce locality by working with quantities defined at each point. These are the fields corresponding to the various physical quantities under consideration.  The laws of physics are expressed by demanding that these fields satisfy field equations.  This takes us over steps four and five.

The sixth step concerns a particular way of defining fields - namely using using local tensor fields based on tangent space.  It is worth elaborating a little here to point out that these are not the most general field theoretic objects we might define on a manifold.  Consider the following
\begin{enumerate}
\item  The tensor fields used in General Relativity can be expressed with respect to a basis choice for each index, $\mu$, at each point, $p\in\mathcal{M}$. The basis choice used corresponds to the (flat) tangent space which approximates the manifold at this point.  This is achieved by using partial derivative operators, $\partial_\mu$, to span the tangent space. However, we could imagine setting up a local basis corresponding to a curved (rather than flat) approximation to the manifold at each point.  We would do this by using higher derivatives for our basis \cite{kolar1999natural}.
\item The tensor fields used in General Relativity are local: they are defined at each point, $p\in\mathcal{M}$.  One could imagine more general objects that were defined at two points, $p,q\in\mathcal{M}$ or, indeed, defined at even more points, or even a continuum of points.  In quantum theory, entangled states can be thought of as corresponding to such multipoint tensors.  Even in classical probability theory, correlations can be represented by multipoint tensors \cite{hardy2010formalism}.
\end{enumerate}

Following these first six steps we have a pretty clear picture emerging.  Physics written in terms of these local tensor fields based on tangent spaces.  Equations in terms of such tensor fields give local physics in the weak sense that they specify relationships between physical quantities at each point $p\in\mathcal{M}$.  As an aside, it is worth mentioning that General Relativity also has a stronger locality property (we will call this \emph{signal locality}) in that it is not possible to signal faster than the speed of light.  It inherits the signal locality property from Special Relativity. There is, however, more to the matter than this as we need to be sure that the non-linear equations of General Relativity do not introduce some faster-than-light signalling possibilities \cite{geroch1972general}. In the theory of partial differential equations, this has to do with the properties of hyperbolic partial differential equations.

There is, in general, no global inertial reference frame. Hence, there is no special choice of coordinate system - all coordinate systems are equally good.  It is reasonable, then, to demand that the equations of the theory take the same form in any coordinate system.  This (the seventh step) is the principle of general covariance:
\begin{quote}
\textbf{The principle of general covariance}: The laws of physics should be written in such a way that they work for any coordinate system and so that they take the same form in any coordinate system.
\end{quote}
This is the seventh step and puts in place the mathematical and conceptual framework within the theory of General Relativity is formulated.

\section{The three elements of General Relativity}\label{sec:elementsofGR}

Having arrived at a general framework following the above seven steps now I want to provide the extra ingredients that specify General Relativity.  I do this by giving three elements.  These provide a statement of what General Relativity is within this general framework.

Before this, however, we need to define what a diffeomorphism is.  A diffeomorphism is a smooth invertible map, $\varphi(p)$, that maps each point $p$ on a manifold $\mathcal{M}$ to a new point, $\varphi(p)$.  If we cover the manifold with coordinates, then a diffeomorphism corresponds to an active coordinate transformation (a diffeomorphism is, then, the abstract version of a general coordinate transformation).

The three elements of General Relativity consist of a prescription, an addendum, and an interpretation:
\begin{description}
\item[A prescription] to convert special relativistic matter field equations into diffeomorphism invariant general relativistic matter field equations by taking the following steps
\[ \eta_{\bar{\mu}\bar{\nu}} \rightarrow g_{\mu\nu} \]
\[\partial_{\bar{\mu}} \rightarrow \nabla_\mu\]
\[x^{\bar{\mu}} \rightarrow x^\mu \]
This technique is often called called \emph{minimal substitution}. Here the indices $\bar{\mu}$ correspond to a global inertial reference frame while the indices $\mu$ correspond to a general coordinate system. Note that the term \lq\lq matter" refers to all physical fields (electromagnetic, fluid, \dots) other than the gravitational field, $g_{\mu\nu}$.
\item[An addendum.] We now have an extra ten real variables in the symmetric tensor, $g_{\mu\nu}$, so we need an extra ten equations
\[ G^{\mu\nu} = 8\pi T^{\mu\nu} \]
These are the Einstein field equations (also diffeomorphism invariant). There are ten of them since $G^{\mu\nu}$ and $T^{\mu\nu}$ are symmetric.  All well and good it would seem.  There is, however, a surprising twist in the tale.  It follows from the mathematical identity, $\nabla_\mu G^{\mu\nu}=0$, that only six of the ten Einstein field equations are independent.  This is not a problem because of the last element \dots
\item[An interpretation.] A solution is given by specifying the fields $\pmb{\Phi}$ (consisting of matter fields and the metric) at each point, $p$, on a manifold, $\mathcal{M}$
\[ \Psi = \left\{ (\pmb{\Phi}, p): \forall p\in\mathcal{M} \right\}  \]
where this must satisfy the field equations (matter plus Einstein) at every point $p$.  The interpretation is that beables are those properties of solutions that are invariant under general diffeomorphisms. Thus, $B$ is a beable if and only if
\[ B(\Psi) = B(\varphi^*\Psi) ~~~~~\forall ~\varphi  \]
where $\varphi^*\Psi$ is the solution obtained by applying a diffeomorphism to $\Psi$ (see \cite{hardy2016operational}).
\end{description}
Here I have adopted John Bell's term, \emph{beables}, \cite{bell1985exchange} for the properties that are taken to be ontologically real.  The minimal substitution technique fails to provide a unique prescription when we have higher than first derivatives (as partial derivatives commute while covariant derivatives do not).  Maxwell's equations and perfect fluids involve only first derivatives so there is a substantial part of physics where this ambiguity is not an issue.

The need for the first element follows from the equivalence principle for two reasons: (i) because it tells us we have to go to general coordinates as outlined above and (ii) also because it tells us that there is a local inertial reference frame at every point (a freely falling one) where we have Special Relativistic physics locally - this is ensured by using the covariant derivative which is equal to the partial derivative in a local inertial reference frame.  The need for the second element follows from the first element because we need extra field equations (beyond the matter field equations) so we can solve for the ten real fields in the metric.  The third element follows from the second as Einstein's field equations only provide six equations (we will discuss the interpretation in more detail in Sec.\ \ref{sec:GRsuggests}.

We can see very clearly how General Relativity solves the problem of Relativistic Gravity in the these elements. In particular, note that the first element says that the special relativistic matter field equations are changed when we go to General Relativity but in such a way that we recover Special Relativity when the effects of gravity can be ignored (when spacetime is flat and there exists a global inertial reference frame).  The second element shows us how Newton's law of gravity is changed.  Einstein's field equations give Newton's law of universal gravitation (in the form of Poison's equation for the potential) in the limit of weak gravitational fields.  Thus, General Relativity is even handed with respect to the two less fundamental theories.  We should, I anticipate, expect the same sort of evenhandedness for Quantum Gravity.

\section{How GR suggests its own interpretation}\label{sec:GRsuggests}

It is important not to let the Einstein field equations, beautiful as they are, distract us from the bigger picture.  The bigger picture is that provided above: first we need the framework then we have the prescription, addendum (Einstein's field equations), and interpretation.  This bigger picture amounts to a radical new view of reality.   Let us look now at the interpretation.  Even after providing Einstein's field equations, we have four fewer field equations than real fields. Further, these field equations are all invariant under diffeomorphisms. These two facts go hand in hand because a diffeomorphism is described by four equations.  This is clear if we return to using general coordinates. Then we have four equations $x^{\mu'} = f^{\mu'}(x^\mu)$ describing the transformation. We can find solutions, $\Psi$, for the field equations.  Further, if we act on any solution with a diffeomorphism, we will find another solution, $\varphi^*\Psi$, that is also consistent with the field equations (since these field equations are invariant under diffeomorphisms). We can argue that these two situations are physically equivalent as follows (this is Einstein's hole argument \cite{stachel2014hole, sep-spacetime-holearg}).  First, consider a compact region, $\mathcal{H}\subset \mathcal{M}$, for which we want to solve the field equations. Assume that, everywhere outside this region, we know all the fields.  Thus, we should have sufficient boundary information to determine the physical situation inside $\mathcal{H}$.  However, for any solution, $\Psi$, we find (consistent with this boundary information) there exists an infinity of solutions $\varphi^*\Psi$ where the diffeomorphism, $\varphi$, is different from the identity only in $\mathcal{H}$. These solutions have the same fields outside $\mathcal{H}$. This means that no matter how much boundary information we provide for $\mathcal{H}$ there are still an infinity of solutions to  the field equations differing inside $\mathcal{H}$.  This leaves us with two choices. Either there is a breakdown of determinism (in the strong sense that boundary conditions fail, even probabilistically, to determine physical properties) or these different solutions correspond to the same physical situation.  We have to choose the latter if we want to have a sensible physical theory.  Thus, General Relativity suggests its own ontological interpretation.  It is worth noting that these diffeomorphisms leave unchanged certain key properties such as whether two fluid blobs intersect or not.  Diffeomorphisms also leave unchanged the distance between pairs of such blob intersection events.  The familiar world of tables, chairs and so on lives amongst the beables of the theory since such objects can be understood in terms of scalars (such as matter densities) taking values in coincidence along with distances between such coincidences.


\part{Quantum Field Theory}\label{part:QFT}

\section{Preliminaries}

If we take the limit of Quantum Gravity to a case where spacetime is fixed, then we expect to obtain some form of Quantum Field Theory.  Now QFT is a very big subject which I could not hope to cover in this paper.  Rather, I will give my own take on the subject in the form of what I call Operator Tensor Quantum Field Theory \cite{hardy2016operational}.  This approach brings out the operational aspects and is being developed with the kind of program discussed in this paper in mind.

I originally proposed the idea for operator tensor QFT in 2011 (see Sec.\ 16 of \cite{hardy2011reformulating}) and developed the idea in \cite{hardy2016operational}.   The closest other approach to this is that of Oeckl.  He initiated his pioneering general boundary formalism in 2003 \cite{oeckl2003general}. This was originally seen as a way to formulate Quantum Theory, Quantum Field Theory, and Quantum Gravity.  In its original incarnation Oeckl's formalism puts states (represented as elements of a Hilbert space) on the boundaries of arbitrary regions of space-time.  This is a pure state formalism linear in quantum amplitudes.  In 2012, motivated by work in the GPT approach, he developed the \emph{positive formalism} \cite{oeckl2013positive} in which positive operators are associated with regions of spacetime.  This, like the operator tensor approach, is linear in probabilities (more recently Oeckl has developed an even broader GPT point of view \cite{oeckl2016local}).  Though the details differ, there is much similarity in spirit between the positive formalism and the operator tensor approach to Quantum Field Theory.

\section{Operator tensor Quantum Theory}\label{sec:optensQT}

There is not agreement as to the correct ontological interpretation of Quantum Theory.  However, there is broad agreement on the operational interpretation.  Here I will summarize the operator tensor formulation of Quantum Theory \cite{hardy2011reformulating, hardy2012operator} as a prelude to doing the QFT case.

In this approach, an operation is represented by a box with some number of quantum systems going in the bottom and some number coming out the top (so we can think of time as flowing up the page)
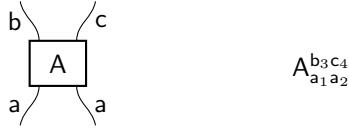
\[
\begin{Compose}{0}{0}\setdefaultfont{\mathsf}\setsecondfont{\mathsf}
\Crectangle{A}{1.5}{1.2}{0,0} \thispoint{Ald}{-2,-3.3} \thispoint{Ard}{2,-3.3} \thispoint{Alu}{-2,3.3} \thispoint{Aru}{2,3.3}
\joincbnoarrow[left]{Ald}{80}{A}{-1} \csymbolalt[-10,0]{a}
\joincbnoarrow[right]{Ard}{100}{A}{1} \csymbolalt[10,0]{a}
\jointcnoarrow[left]{A}{-1}{Alu}{-80} \csymbolalt[-10,0]{b}
\jointcnoarrow[right]{A}{1}{Aru}{-100} \csymbolalt[10,0]{c}
\end{Compose}
~~~ ~~~~~ ~~~ ~~~~~~~~~
\mathsf{A}_{\mathsf{a}_1\mathsf{a}_2}^{\mathsf{b}_3\mathsf{c}_4}
\]
We can also represent an operation symbolically as shown on the right. Then subscripts represent inputs and superscripts represent outputs. Different types of quantum system (electrons, photons, protons, \dots) are represented by different labels, $\mathsf{a}$, $\mathsf{b}$, $\mathsf{c}$, \dots.  An operation will have a setting, $Q_\mathsf{A}$, corresponding to the positions knobs and suchlike on the associated apparatus.  This apparatus will also have outcomes (corresponding to different lights flashing).  Associated with a given operation will be some given set of outcomes, $O_\mathsf{A}$.  A \emph{complete set of operations}
\[
\{\mathsf{A}_{\mathsf{a}_1\mathsf{a}_2}^{\mathsf{b}_3\mathsf{c}_4}[i]: i=1 ~\text{to}~ L_\mathsf{A}\}
\]
is a set having a given setting and whose outcome sets $O_\mathsf{A}[i]$ are disjoint with union equal to the set of all possible outcomes on this apparatus.

We can wire operations together to form circuits (if no open wires are left over) or fragments (of circuits) if some are.  For example
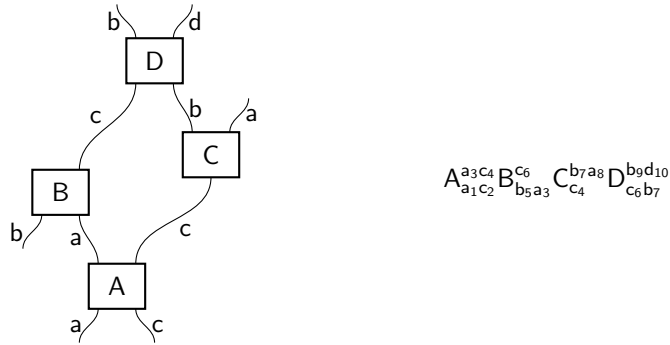
\[
\begin{Compose}{0}{-1.5}\setdefaultfont{\mathsf}\setsecondfont{\mathsf}
\thispoint{DUL}{0,15}\thispoint{DUR}{4, 15}
  \Crectangle{D}{1.5}{1.2}{2,12}
                                                  \thispoint{CUR}{7,10}
\Crectangle{B}{1.5}{1.2}{-3, 5}   \Crectangle{C}{1.5}{1.2}{5,7}
\thispoint{BDL}{-5,2}

      \Crectangle{A}{1.5}{1.2}{0,0}
\thispoint{ADL}{-2,-3}     \thispoint{ADR}{2,-3}
\jointbnoarrow[left]{D}{-1}{DUL}{0} \csymbolalt[-5,0]{b} \jointbnoarrow[right]{D}{1}{DUR}{0}\csymbolalt[5,0]{d}
\jointbnoarrow[above left]{B}{1}{D}{-1} \csymbolalt[-4,0]{c} \jointbnoarrow[right]{C}{-1}{D}{1} \csymbolalt[5,0]{b} \jointbnoarrow[right]{C}{1}{CUR}{0} \csymbolalt[5,0]{a}
\jointbnoarrow[left]{BDL}{0}{B}{-1} \csymbolalt[-10,0]{b}\jointbnoarrow[left]{A}{-1}{B}{1} \csymbolalt[-5,0]{a} \jointbnoarrow[below right]{A}{1}{C}{0} \csymbolalt[4,0]{c}
\jointbnoarrow[left]{ADL}{0}{A}{-1} \csymbolalt[-5,0]{a} \jointbnoarrow[right]{ADR}{0}{A}{1} \csymbolalt[5,0]{c}
\end{Compose}
~~~~~~~~~~~~~~~~~~~~
\mathsf{A}_\mathsf{a_1c_2}^\mathsf{a_3c_4} \mathsf{B}_\mathsf{b_5a_3}^\mathsf{c_6} \mathsf{C}_\mathsf{c_4}^\mathsf{b_7a_8} \mathsf{D}_\mathsf{c_6b_7}^\mathsf{b_9d_{10}}
\]
is a fragment.  In the symbolic notation we use repeated subscripts (one raised one lowered) with an integer label for the given wire.

Associated with an operation is an Hermitian operator (which we can represent diagrammatically or symbolically).
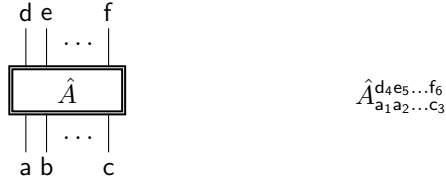
\[
\begin{Compose}{0}{0}\setdefaultfont{\hat}\setsecondfont{\mathsf}
\Crectangledouble{A}{3}{1.2}{0,0}
\thispoint{a1}{-2.2,-3.3}  \csymbolalt[0,-25.5]{a}    \thispoint{b2}{-1.1,-3.3} \csymbolalt[0,-21]{b} \Thistexthere{\dots}{0.7,-2.5} \thispoint {c3}{2.2,-3.3} \csymbolalt[0,-25]{c}
\jointbnoarrow{a1}{0}{A}{-2.2}                      \jointbnoarrow{b2}{0}{A}{-1.1}                                                 \jointbnoarrow{c3}{0}{A}{2.2}
\thispoint{d4}{-2.2,3.3}  \csymbolalt[0,25]{d}    \thispoint{e5}{-1.1,3.3} \csymbolalt[0,22]{e} \Thistexthere{\dots}{0.7,2.5} \thispoint {f6}{2.2,3.3} \csymbolalt[0,25]{f}
\joinbtnoarrow{d4}{0}{A}{-2.2}                      \joinbtnoarrow{e5}{0}{A}{-1.1}                                                 \joinbtnoarrow{f6}{0}{A}{2.2}
\end{Compose}
~~~~~~~~~~~~~~~~~~~~~~~~~~
\hat{A}_\mathsf{a_1 a_2\dots c_3}^\mathsf{d_4 e_5 \dots f_6}
\]
This operator acts on a Hilbert space
\[ \mathcal{H}_\mathsf{a_1 b_2\dots c_3}^\mathsf{d_4 e_5 \dots f_6} =
\mathcal{H}_\mathsf{a_1}\otimes\mathcal{H}_\mathsf{b_2}\otimes \dots \otimes \mathcal{H}_\mathsf{c_3} \otimes \mathcal{H}^\mathsf{d_4} \otimes \mathcal{H}^\mathsf{e_5} \otimes \dots \otimes \mathcal{H}^\mathsf{f_6}\]
as determined by the subscripts and superscripts.

We can wire together operators.  Then wires correspond to the partial trace over the appropriate part of the Hilbert space.  To understand this, consider the three expressions
\[ \hat{A}^\mathsf{a_1} \hat{B}_\mathsf{a_1} ~~~~~~ \hat{A}^\mathsf{a_1} \hat{B}_\mathsf{a_1} \hat{C}^\mathsf{b_2} ~~~~~~ \hat{A}^\mathsf{a_1} \hat{D}_\mathsf{a_1}^\mathsf{b_2} \]
The first expression is equal to the trace of the product of the two given operators (a real number).  The second expression is equal to the trace of the product of the first two operators (a real number) times the third operator.  To evaluate the third expression can be evaluated by expanding $\hat{D}_\mathsf{a_1}^\mathsf{b_2}$ out as a sum of products (like $\hat{B}_\mathsf{a_1} \hat{C}^\mathsf{b_2}$) then evaluating for each term.  If we have a circuit (with no open wires) then we obtain a real number.

We will say that operations \emph{correspond} to operators under some mapping if, under this mapping, the probability for \emph{any} circuit comprised of operations is equal to the real number obtained by replacing these operations in this circuit with operators under this mapping.

There are constraints on operators in order that (i) probabilities are between 0 and 1 and (ii) causality is satisfied (so that it is not possible to communicate backward in time).  These \emph{physicality} \cite{hardy2011reformulating, hardy2012operator, hardy2016operational} constraints are
\begin{equation}\label{QTphysicality}
0 \leq  \hat{A}_\mathsf{a_1^T a_2^T}^\mathsf{b_3 c_4} ~~~~
\hat{A}_\mathsf{a_1 a_2}^\mathsf{b_3 c_4} \hat{I}_\mathsf{b_3 c_4} \leq \hat{I}_\mathsf{a_1 a_2}
\end{equation}
where $\mathsf{T}$ denotes taking the partial transpose over the associated part of the Hilbert space and $\hat{I}_\mathsf{a_1 a_2}$ is the identity operator acting on $\mathcal{H}_\mathsf{a_1 a_2}$ (these physicality constraints relate to the Pavia causality conditions \cite{chiribella2010probabilistic}).  Note that the second of these physicality conditions is time asymmetric. A complete set of physical operators,
\[
\{\mathsf{A}_{\mathsf{a}_1\mathsf{a}_2}^{\mathsf{b}_3\mathsf{c}_4}[i]: i=1 ~\text{to}~ L_\mathsf{A}\}
\]
is a set of physical operators satisfying the normalization condition
\[   \sum_i \hat{A}_\mathsf{a_1 a_2}^\mathsf{b_3 c_4}[i] \hat{I}_\mathsf{b_3 c_4} = \hat{I}_\mathsf{a_1 a_2}   \]
We can now give the following simple axiom for Quantum Theory
\begin{quote}
\textbf{Axiom for QT:} All complete sets of operations correspond to complete sets of physical operators and vice versa.
\end{quote}
This succinct statement captures the content of Quantum Theory.  In particular, as shown in \cite{hardy2011reformulating, hardy2012operator}, it follows that preparations are associated with positive operators with trace less than or equal to one (density matrices), evolution is associated with completely positive trace non-increasing maps, and measurement is associated with positive operator valued measures.

\section{Operator Tensor Quantum Field Theory}\label{sec:optensQFT}

First consider the following discrete case. We have qubits traveling to the left and to the right (time goes up the page).
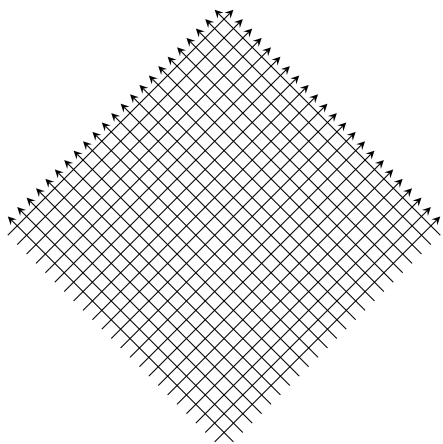
\[
\begin{Compose}{0}{0}
\Cgrid{0.5}{11}{11}{0,0}
\end{Compose}
\]
At each vertex, $p$, in this diagram we assume we have an operation (as discussed above) having setting $Q(p)$ and outcome $O(p)$ (where $p$ labels the vertices).  Now consider a region, $\mathtt{A}$ (consisting of a collection of vertices) such as
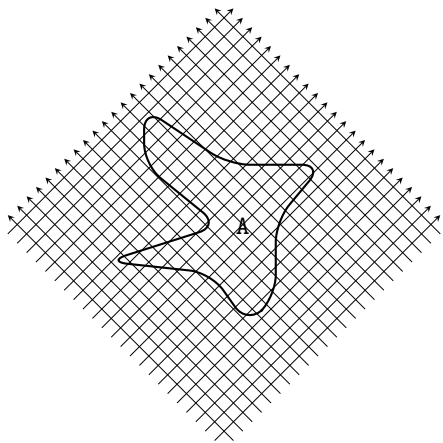
\begin{equation}\label{discreteobject}
\begin{Compose}{0}{0}
\setdefaultfont{\mathtt}
\Cobject{\thetooth}{A}{1}{1}{1,-0.07}
\Cgrid[->, ultra thin]{0.5}{11}{11}{0,0}
\end{Compose}
\end{equation}
If we let the distance between the qubits tend to zero (while keeping $\mathtt{A}$ fixed) then the qubit trajectories become denser and denser inside this region.  As we go over to the continuous case the vertices are labeled by $p\in\mathtt{M}$ where $\mathtt{M}$ is a manifold (or at least, part of a manifold covering the points of interest). For the example shown this will be a $2$ dimensional manifold but we can construct grids that limit to more general $D$ dimensional manifolds.   In this continuous limit the setting field, $Q(p)$, and the outcome field, $O(p)$, become continuous fields.

We now have the following elements.

\paragraph{Manifold.} We have a given manifold (or region of a manifold), $\mathtt{M}$, covering the spacetime points, $p$, of interest.

\paragraph{Metric and time direction fields.} We assume that the manifold is covered by a fixed metric, $g_{\mu\nu}(p)$, and also a fixed time direction field, $\tau^\mu(p)$, that points in the forward light cone at each point $p$.  We assume there are no closed forward pointing timelike paths in $\mathtt{M}$ with respect to $g_{\mu\nu}$ and $\tau^\mu$. We need the time direction field as the physicality conditions are time asymmetric.  There is a certain gauge freedom in defining the time direction field.  The field ${\tau'}^\mu(p)$ is physically equivalent to $\tau^\mu(p)$ if it points into the same light cone at each point $p$.

\paragraph{Typing surfaces.}  A typing surface is represented by
\[ \mathtt{a} = \{ (p, n^\mu): \forall p\in \text{set}(\mathtt{a}) \}    \]
where $\text{set}(\mathtt{a})$ is a surface (having dimension one less than $\mathtt{M}$) and  $n^\mu(p)$ is a vector that is not anywhere embedded in the surface (it's job is to single out one side or the other of this surface).  Typing surfaces are used to bound all or part of a region such as $\mathtt{A}$. The direction is a conventional choice.  As is clear from considering the discrete situation shown in \eqref{discreteobject} above, sections of the boundary may correspond to quantum systems being inputted, outputted, or a mixture of both.  We define
\[ \mathtt{a^R} = \{ (p, -n^\mu): \forall p\in \text{set}(\mathtt{a}) \}    \]
We use the notation
\[ (\mathtt{c,ab}) = \mathtt{c\cup a^R \cup b^R }  \]
for composite typing surfaces. We allow these typing surfaces to meet at their boundaries.  Consequently, some points $p$ may appear twice with different values of $n^\mu(p)$.

\paragraph{Setting and outcome fields.}  We can chose settings, $Q(p)$, at point $p$ and have outcome, $O(p)$, at $p$.

\paragraph{Operations.}  An operation, $\mathsf{A}_\mathtt{ab}^\mathsf{c}$, is specified with respect to some region, $\mathtt{A}\in\mathtt{M}$, as follows
\[ \mathsf{A}_\mathtt{ab}^\mathtt{c}  = \{ Q_\mathsf{A}, O_\mathsf{A}, \mathtt{A}, (\mathtt{c,ab}) \} \]
where
\[ Q_\mathtt{A} = \{ (p, Q(p)): \forall p\in\mathtt{A} \}   \]
and
\[ O_\mathtt{A} = \{ (p, O(p)): \forall p\in\mathtt{A} \}   \]
and the typing surface is $(\mathtt{c,ab})$. The typing surface may bound all of $\mathtt{A}$ or it may bound only part of $\mathtt{A}$ (for example, we could have absorbing detectors or preparation devices included in the part of the boundary not covered by the typing surface).
We can also represent operations diagrammatically
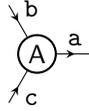
\[
\begin{Compose}{0}{0}\setdefaultfont{\mathsf}\setsecondfont{\mathtt}
\Ucircle{A}{0,0}  \thispoint{a}{0:3} \thispoint{b}{120:3} \thispoint{c}{-120:3}
\joincc[above]{A}{0}{a}{-180} \csymbolalt{a} \joincc[above right]{b}{-60}{A}{120} \csymbolalt{b} \joincc[below right]{c}{60}{A}{-120} \csymbolalt{c}
\end{Compose}
\]
We use a circle rather than a rectangle to represent these operations to emphasize that they are different from the operations considered in Sec.\ \ref{sec:optensQT}. Here an arrow pointing out is conventional and does not have to correspond to systems progressing forward in time.
We say that
\[ \{ \mathsf{A}_\mathtt{ab}^\mathsf{c}[i]: i=1 ~\text{to}~L_\mathtt{A} \}  \]
constitutes a \emph{complete set of operations} if the associated outcome sets are disjoint and their union is the set of all possible outcomes for region $\mathtt{A}$.

\paragraph{Operators.} We can associate operator tensors with a region $\mathtt{A}$.  To do this, first we associate Hilbert spaces with typing surfaces
\[
\mathcal{H}^\mathtt{a} = \mathcal{H}^\mathtt{a_+}\otimes \mathcal{H}^\mathtt{a_-}
~~~~~~
\mathcal{H}_\mathtt{a} = \mathcal{H}_\mathtt{a_+}\otimes \mathcal{H}_\mathtt{a_-}
\]
where the $+$($-$) will pertain to a time direction.  An operator tensor is given by (in diagrammatic and symbolic notation)
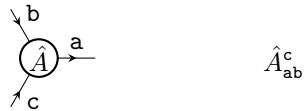
\[
\begin{Compose}{0}{0}\setdefaultfont{\hat}\setsecondfont{\mathtt}
\Ucircle{A}{0,0}  \thispoint{a}{0:3} \thispoint{b}{120:3} \thispoint{c}{-120:3}
\joincc[above]{A}{0}{a}{-180} \csymbolalt{a} \joincc[above right]{b}{-60}{A}{120} \csymbolalt{b} \joincc[below right]{c}{60}{A}{-120} \csymbolalt{c}
\end{Compose}
~~~~~~~~~~~~~~~~~~~
\hat{A}_\mathtt{ab}^\mathtt{c}
\]
and is an element of $\mathcal{V}_\mathtt{ab}^\mathtt{c}$
- the space of Hermitian operators acting on $\mathcal{H}_\mathtt{a} \otimes \mathcal{H}_\mathtt{b} \otimes \mathcal{H}^\mathtt{c}$.

\paragraph{Composition.}  We can wire operations together to form bigger operations.  For example
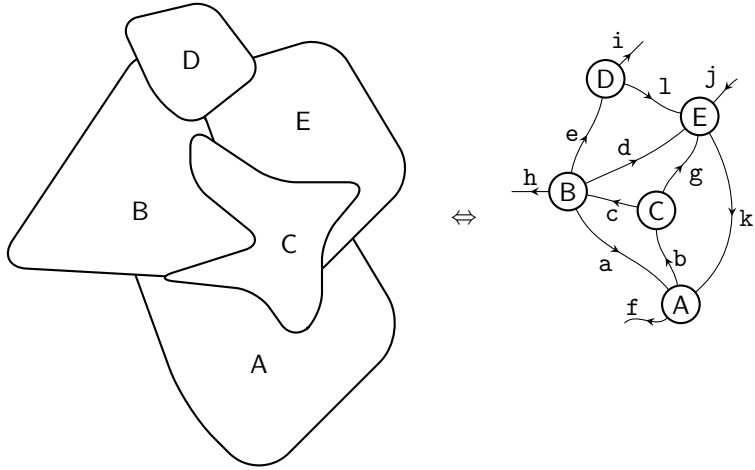
\[
\begin{Compose}{0}{0}
\cobjectwhite{\theblob}{A}{2}{2}{-0.2,-5} \csymbol[0,-90]{A}
\cobjectwhite{\theflag}{B}{2}{1}{-4.75,0} \csymbol[-50,0]{B}
\cobjectwhite{\theblob}{E}{1.5}{1.5}{2.15,3} \csymbol[0,50]{E}
\cobjectwhite{\thetooth}{C}{1}{1}{1.4,-1.75} \csymbol{C}
\cobjectwhite{\theblob}{D}{0.95}{0.75}{-3.8,8} \csymbol{D}
\end{Compose}
~~~ \Leftrightarrow ~~~
\begin{Compose}{0}{-1} \setdefaultfont{\mathsf}\setsecondfont{\mathtt}
\Ucircle{A}{1,-1} \Ucircle{B}{-5,5} \Ucircle{C}{-0.3,4} \Ucircle{D}{-3, 11} \Ucircle{E}{2,9}
\joincc[below left]{B}{-65}{A}{130} \csymbolalt{a}
\joincc[right]{A}{100}{C}{-90} \csymbolalt{b}
\joincc[below]{C}{170}{B}{-10} \csymbolalt{c}
\joincc[above left]{B}{25}{E}{-140} \csymbolalt{d}
\joincc[left]{B}{80}{D}{-100} \csymbolalt{e}
\joincc[above right]{D}{-15}{E}{170}\csymbolalt{l}
\joincc[below right]{C}{70}{E}{-95} \csymbolalt{g}
\joincc[right]{E}{-60}{A}{40}\csymbolalt{k}
\thispoint{nA}{-2,-2} \joincc[above left]{A}{-135}{nA}{45} \csymbolalt{f}
\thispoint{nB}{-8,5} \joincc[above]{B}{180}{nB}{0} \csymbolalt{h}
\thispoint{nD}{-1,13} \joincc[above left]{D}{45}{nD}{-135} \csymbolalt{i}
\thispoint{nE}{4,11} \joincc[above left]{nE}{-145}{E}{45} \csymbolalt{j}
\end{Compose}
\]
If there are no open wires then we have a circuit.

\paragraph{Physicality.}  We require, when we replace operations by operators in a circuit, that we get a number between 0 and 1 for the probability.  We also require that we cannot signal into the past.  These two requirements are satisfied if certain physicality properties are satisfied.  These physicality conditions are more complicated than for the operations (represented by rectangles) of Quantum Theory as discussed in Sec.\ \ref{sec:optensQT} for two reasons. First, the causal structure for an operation in QFT (represented by a circle) is different. In particular, for a QFT operation, it is possible for an output to be fed forward back into the same operation.  This can be made clear by looking at the discrete situation in \eqref{discreteobject}.   Second, we have a continuous rather than discrete situation.  We will say that an operator, $\hat{A}_\mathtt{ab}^\mathtt{c}$, pertaining to region $\mathtt{A}$ is physical if it satisfies the condition
\[ \text{Physicality}_\text{QFT}(\hat{A}_\mathtt{ab}^\mathtt{c}, \mathtt{A})   \]
I will describe this condition in some detail in the appendix (though there still remains more work to be done in understanding how physicality plays out in QFT).
We say that
\[ \{ \hat{A}_\mathtt{ab}^\mathsf{c}[i]: i=1 ~\text{to}~L_\mathtt{A} \}  \]
represent a \emph{complete set of physical operators} if each operator in the set is physical and they satisfy the normalization condition
\[   \sum_i \hat{A}_\mathsf{a_1 a_2}^\mathsf{b_3 c_4}[i] \hat{I}_\mathsf{b_3 c_4} = \hat{I}_\mathsf{a_1 a_2}   \]
(see \cite{hardy2016operational} for clarification).

\paragraph{Quantum Field Theory.} We can now state an axiom for QFT
\begin{quote}
\textbf{Axiom for QFT:} All complete sets of operations correspond to complete sets of physical operators and vice versa.
\end{quote}
This succinct statement captures the essence of Quantum Field Theory in the operator tensor formulation.


\part{A conceptual road to Quantum Gravity}\label{part:roadtoQG}

\section{Introduction}

I will now outline one approach to finding a theory of Quantum Gravity.  Before I begin, I should emphasize once again that The Construction Interpretation is pluralistic in that it explicitly encourages a multitude of different conceptually driven approaches to the problem of Quantum Gravity in accord with the tenets given above.  There is value, however, in outlining a particular approach if only to illustrate the general ideas of the Construction Interpretation.

\section{The problem of Quantum Gravity}

As discussed in the prelude, the problem of Quantum Gravity is to find a theory that gives rise to General Relativity on the one hand and to Quantum Field Theory on the other under appropriate limits.
\[
\text{General Relativity} \longleftarrow \text{Quantum Gravity} \longrightarrow \begin{array}{l} \text{Special Relativistic} \\  ~~~\text{Quantum Field Theories} \end{array}
\]
At least we must recover the parts of these theories that have been empirically verified.  While a theory of Quantum Gravity must have these limits, the conceptual and mathematical formalism could be very different from either General Relativity or Quantum Theory.  Indeed, a theory of Quantum Gravity could be as different from either General Relativity or Quantum Theory as, for example, General Relativity is from Newtonian Mechanics.  There may be more than one candidate theory of Quantum Gravity that has these limits. If so, we might try to distinguish between these alternative theories on the grounds of simplicity, elegance, or conceptual coherence. In the end, of course, experiment is the final arbitrator.  A theory of Quantum Gravity would be especially interesting if it has features that are genuinely new and cannot be understood in solely General Relativistic or Quantum terms.

\section{Path to a framework for Quantum Gravity}\label{sec:pathtoaframework}

We can attempt to plot a path with $7$ steps $+$ $3$ elements for Quantum Gravity which is analogous to the path for General Relativity outlined earlier. In this path indefinite causal structure (inferred from some facts about General Relativity and Quantum Theory as we will outline) plays an analogous role to the fact that there is no global inertial reference frame (inferred from some facts about falling bodies) in the General Relativity path outlined above.  The path proposed here is \emph{analogous} to that for General Relativity outlined earlier rather than being the \emph{same} path.

In brief the path I propose consists, first, of the following 7 steps to bring us to a general mathematical framework for Quantum Gravity
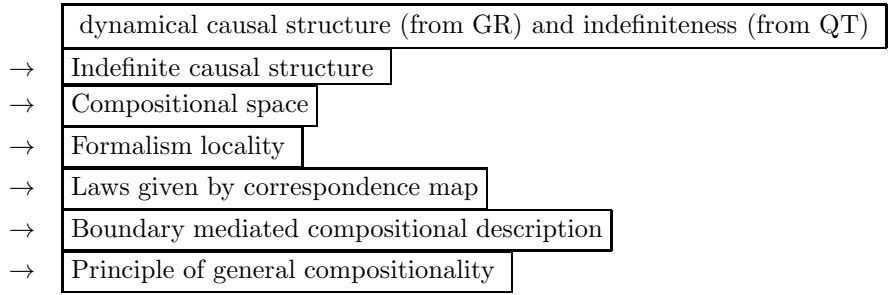
\[
\begin{array}{ll}
{}          & \fbox{ \text{dynamical causal structure (from GR) and indefiniteness (from QT)} } \\
\rightarrow & \fbox{\text{Indefinite causal structure} }\\
\rightarrow &\fbox{\text{Compositional space}} \\
\rightarrow &\fbox{\text{Formalism locality} } \\
\rightarrow &\fbox{\text{Laws given by correspondence map}} \\
\rightarrow &\fbox{\text{Boundary mediated compositional description}} \\
\rightarrow &\fbox{\text{Principle of general compositionality} }
\end{array}
\]
Proceeding by analogy with Einstein's path to General Relativity, we can write down 3 additional steps that might help us get to an actual theory of Quantum Gravity formulated within this framework. These are
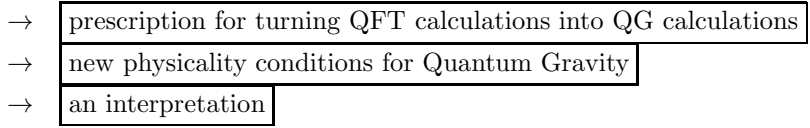
\[
\begin{array}{ll}
\rightarrow &\fbox{\text{prescription for turning QFT calculations into QG calculations}} \\
\rightarrow &\fbox{\text{new physicality conditions for Quantum Gravity}} \\
\rightarrow &\fbox{\text{an interpretation}}
\end{array}
\]
I will now outline in some detail the $7+3$ steps in this path insofar as they are clear to me (the last three steps being particularly opaque).  We will indicate which element is being considered at the beginning of the relevant section along with which element this is analogous to in the path to General Relativity outlined in Sec.\ \ref{sec:EinsteinPath}

It is worth mentioning at the outset that indefinite causal structure raises big challenges for our usual way of doing physics. In particular, we cannot consider a state on a spacelike hypersurface and then evolve this in time.  We need a different way to do physics.

\section{Step one: Some features of GR and QT}

\begin{center}\fbox{ \text{Dynamical causal structure (from GR) and indefiniteness (from QT)} } \end{center}

\begin{center}analogous to \end{center}

\begin{center}\fbox{\text{Equivalence principle} }\end{center}

\vskip 3mm

\noindent
To gain clues as to what kind of theory might limit to GR and QT we should examine the conceptual structure of these less fundamental theories.  In so doing we see that they are each conservative and radical, though in complementary respects as summarized in the following table
\begin{center}
\begin{tabular}{ c| c | c}
   {}                & \text{GR}                          & \text{QT}                    \\
\hline
\text{conservative}& \text{deterministic}               & \text{fixed causal structure}\\
\hline
\text{radical}     & \text{dynamical causal structure}  & \text{probabilistic}
\end{tabular}
\end{center}
General Relativity is conservative in that, as discussed above, if we specify sufficient boundary information the physical situation is determined.  On the other hand, Quantum Theory is radical in that it is inherently probabilistic.  Quantum Theory is conservative in that it operates on a fixed causal background structure.  On the other hand, General Relativity is radical in that the causal structure (given by the metric) is dynamically determined by the equations of the theory.

In Sec.\ \ref{sec:quantumequivprinc} I will discuss an even stronger possible analogy with the equivalence principle - this is the \emph{quantum equivalence principle}: that it is always possible to find a quantum frame of reference in which indefinite causal structure is transformed away locally at any point (but not globally). This principle is more speculative and it is not completely clear, at this stage, how it fits in with the main narrative of these seven steps.

\section{Step two: Indefinite causal structure}\label{sec:steptwo}

\begin{center}\fbox{ Indefinite causal structure}\end{center}
\begin{center} analogous to \end{center}
\begin{center} \fbox{\text{No global inertial reference frame}} \end{center}

\noindent
It seems that a theory of Quantum Gravity would need to take the radical road in both cases. Thus, it would be a probabilistic theory with dynamical causal structure.  But, actually, it would likely be even more radical for the following reason. In Quantum Theory, quantities subject to variation are also subject to quantum indefiniteness (or \lq\lq no-matter-of-the-factness"). If a particle can go through the left or the right slit then there is no-matter-of-the-fact as to which slit it goes through.  If causal structure is dynamical then we expect there to be situations in which there is no matter-of-the-fact as to what the causal structure is. This is indefinite causal structure.  For example it could be the case that, for example, there is no matter-of-the-fact as to whether the separation between some particular two events (identified in operational terms perhaps) is space-like or time-like.



\section{Step three: Compositional space}

\begin{center}\fbox{\text{Compositional space}}\end{center}
\begin{center} analogous to \end{center}
\begin{center} \fbox{\text{General coordinates}} \end{center}

\subsection{Need for a new compositional space}\label{sec:compspace}

The world is a big place and we need a way to break it up (conceptually) into smaller pieces if we are to make sense of it.  We can then join these smaller pieces back together to get the bigger picture. We will call the space we use to do this the \emph{compositional space}. Usually this role is played by spacetime.

A standard way of breaking the world up into such smaller pieces is to evolve (in time) a state defined on a spacelike hypersurface.  We can think of the infinitesimal timesteps as defining small regions of spacetime (narrow in time while being wide in space). This approach is blocked if we have indefinite causal structure since then we cannot, in general, talk about spacelike hypersurfaces.

A more general approach is to break spacetime up arbitrarily into small regions and then have rules at the boundaries of these regions for how the regions fit together.  Indeed, we can think of local field equations (more-or-less) in this way - they relate local derivatives and thus provide constraints on what can happen in an infinitesimal region around any point. This approach is also blocked if we have indefinite causal structure as we will now discuss.

In General Relativity spacetime is represented by points in a manifold.  Consider a region $\mathcal{A}\subset\mathcal{M}$ of this manifold.  There are no beables corresponding to this region as there is no diffeomorphism invariant function $B(\Psi)$ that only looks at the fields in $\mathcal{A}$.  This is because, if we apply a diffeomorphism, then the values for the fields that were in $\mathcal{A}$ will be replaced with values for these fields coming from some other region. There is, then, no non-trivial function of the fields in $\mathcal{A}$ that is invariant under diffeomorphisms. Reality is \lq\lq non-stick" so far as the manifold is concerned. If we want the compositional space to be physical then the manifold is not fit for this role.

We could try to fudge this by introducing a gauge fixing.  This is done by introducing four additional field equations, having no physical content themselves, but providing enough extra field equations to fix the solution.  Not only do these four additional field equations have no physical content, they also obscure the actual physical content of the solution (as it is not clear which features of the solution are a consequence of this gauge fixing and which are real physical features).  Nevertheless, this strategy is mathematically possible and is usually adopted when presenting actual solutions.

The gauge fixing fudge no longer makes sense when we introduce indefinite causal structure.  In fact, even if we have something like a probabilistic ignorance of the classical variety, we can see that gauge fixing is not adequate and we need an alternative compositional space.  In the classical probabilistic case, we will have probabilistic contributions from many different solutions. These different solutions will live on different manifolds, $\mathcal{M}_i$ (where $i$ is labeling the different solutions), and these may even be topologically different.  Then any region, $\mathsf{A}$, of our compositional space would have to correspond to a region, $\mathcal{A}_i\subset \mathcal{M}_i$, of each of the manifolds in this mixture.  Without some principle telling us how the subregions, $\mathcal{A}_i$, correspond for the different $i$, there is no sensible way of doing this.  Indefinite causal structure only makes this worse.  Then we do not have a classical mixture of underlying solutions and it would appear that we are completely blocked from using the usual  manifold  of General Relativity for our compositional space.

\subsection{Beables as axes for compositional space}

A better strategy is, instead, to use some beables to constitute this compositional space.  This could be obtained by taking some nominated (ordered) set of beables,
\[ \mathbf{B}=(B^k: k=1~\text{to}~K) \]
and letting these form the axes of our space.  In a particular run of the experiment only some points in this compositional space would happen (we will illustrate this in the case of General Relativity below).  This is radically different from using a manifold where we have fields at every point.

What considerations can motivate the choice of a subset, $\mathbf{B}$, of beables?   One idea is that they should correspond to our direct experiences of the world.  Thus, we can set up an \emph{operational space} in which the axes are things we directly experience (at some level of description).   Regions of operational space would correspond to regions of the world as identified by our observations.  Such a choice would be contingent on how we look at the world.  Different creatures may nominate different beables for their operational space.  Our experience with Quantum Theory demonstrates that this kind of contingency is not a problem - or at least it is not an obstacle for doing calculations.  In Quantum Theory we can nominate which degrees of freedom correspond to measurement outcomes putting the Heisenberg cut between the classical and quantum worlds wherever is convenient.  As long as sufficient decoherence has occurred, it makes no difference for all practical purposes where we put this split. We should be able to do the same for Quantum Gravity.  This pleasing blending of considerations from General Relativity along with our intuitions from Quantum Theory strongly suggest setting up an operational space for Quantum Gravity.

\subsection{An object in compositional space}

We want to break the world up into smaller pieces in compositional space.  We will call these smaller pieces \emph{objects}.  An object, $\mathsf{A}$, is specified as follows
\[ \text{An object:} ~~~~~ \mathsf{A}= (\text{settings}, \text{happening}, \text{region}, \text{handles})  \]
The \emph{region} refers to a certain region of compositional space.  We will adopt the convention of associating object $\mathsf{A}$ (in \verb|\mathsf| font) with region $\mathtt{A}$ (i.e.\ denoted by the same uppercase letter but in \verb+\mathtt+ font).

The \emph{settings} are things we may choose in region $\mathtt{A}$. These could set by adjusting knobs, or they could correspond to imposing some external field (a magnetic field for example).

The \emph{happenings} specifies some particular thing that may happen.  This may be described at an ontological level or an operational level.  In the operational case we consider only beables that correspond to things we observe.  In this case  we will refer to happenings as \emph{outcomes} (these might correspond to the red light on a box flashing for example).  In the ontological case the happenings could include beables we do not have direct observational access to (traditionally called \lq\lq hidden variables" in the Quantum Foundations literature).

Finally, we have \emph{handles}.  These tell us how this object can be joined to other objects. We will discuss this later.

In the case that our object refers to happenings that are outcomes we will call it an \emph{operation}.
\[  \text{An operation:} ~~~~~ \mathsf{A}= (\text{settings}, \text{outcomes}, \text{region}, \text{handles})  \]
Operational theories allow us to make calculations for composite operations.


\subsection{Operational space and operations in GR}

What kind of beables should we nominate to constitute our operational space?  Here I will discuss the choice of operational space advocated in \cite{hardy2016operational} for General Relativity. This approach borrows from the work of Westman and Sonego \cite{westman2008events, westman2009coordinates} (though they did not have the same operational motivations as here).  Consider an (ordered) set of scalars,
\[\mathbf{S}=(S^1, S^2, \dots S^k)          \]
We can think of a space whose axes are the $S^k$. This is our operational space.  Each scalar, $S^k$, is calculated from the tensor fields, $\pmb{\Phi}$ (these are the matter and metric fields as discussed in Sec. \ref{sec:elementsofGR}), by forming a quantity in which all the indices are summed over.  For example, we might have $S^1=g_{\mu\nu}J^\mu J^\nu$ where $J^\mu$ is a current, $S^2=\rho$ where $\rho$ is a density, and so on.

\begin{figure}
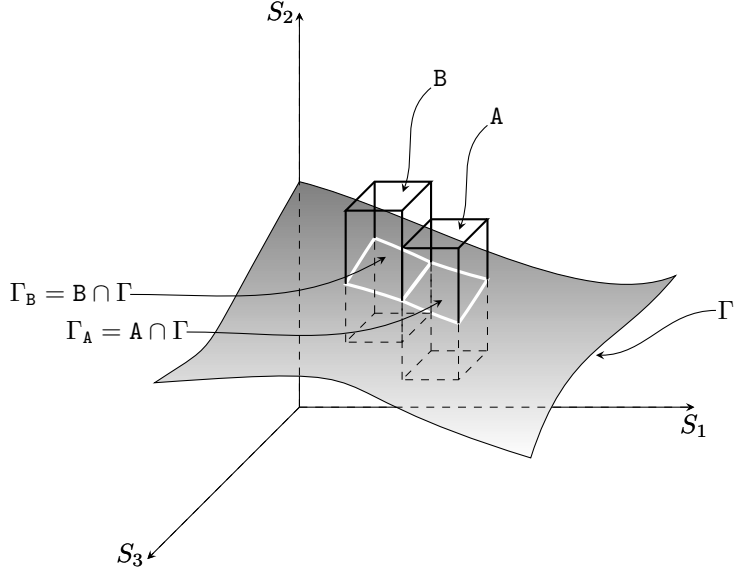

\begin{Compose}{0}{0}
\Cxaxis{S_1}{0,0}{21} \Cyaxis{S_2}{0,0}{21} \Czaxis{S_3}{0,0}{21}
\csurface[0.7]{0,0}{1}
\ccuboid[-0.05]{7,3}{3}{7}{4}{4.7}{3.8}{3}{4.2}
\ccuboid[0.05]{4,5}{3}{7}{4}{4}{2.7}{2.2}{3.1}
\thispoint{GS}{20, 7, 11} \thispoint{G}{26, 9, 10} \pointingarrow{G}{-180}{GS}{0} \csymbolalt[1,0]{\Gamma}
\thispoint{AO}{8.5,9.44} \thispoint{A}{10,15} \pointingarrow[above right]{A}{-140}{AO}{70} \csymbolalt{\mathtt{A}}
\thispoint{BO}{5.5,11.44} \thispoint{B}{7,17} \pointingarrow[above right]{B}{-140}{BO}{70} \csymbolalt{\mathtt{B}}
\thispoint{GAS}{7.6, 5.7} \thispoint{GA}{-6, 4} \pointingarrow[left]{GA}{0}{GAS}{-150} \csymbolalt[-75,0]{\Gamma_\mathtt{A}=\mathtt{A}\cap\Gamma}
\thispoint{GBS}{4.6, 8} \thispoint{GB}{-9, 6} \pointingarrow[left]{GB}{0}{GBS}{-150} \csymbolalt[-75,0]{\Gamma_\mathtt{B}=\mathtt{B}\cap\Gamma}
\Cxaxis[ultra thin, dashed]{S_1}{0,0}{21} \Cyaxis[ultra thin, dashed]{S_2}{0,0}{21} \Czaxis[ultra thin, dashed]{S_3}{0,0}{21}
\end{Compose}
\caption{The surface $\Gamma$ in operational space.  Also shown are two regions $\mathtt{A}$ and $\mathtt{B}$.}\label{fig:AandBcomposed}
\end{figure}

Now consider a solution, $\Psi$, to the field equations of General Relativity (as discussed in Sec.\ \ref{sec:elementsofGR}).
For each $p\in\mathcal{M}$, we can calculate $\mathbf{S}$ from the fields, $\pmb{\Phi}$, at $p$ in this solution. In this way, the point, $p$, gives rise to a point, $\mathbf{S}$, in operational space.  When we repeat this for all points, $p\in\mathcal{M}$, we get a surface, $\Gamma$, in operational space as shown in Fig.\ \ref{fig:AandBcomposed}.
Under a diffeomorphism mapping $p$  to $q=\varphi(p)$, scalars transform as $S^k(p)=S^k(q)$.  Hence, the property of some set of scalars having certain values in coincidence is unaffected by a diffeomorphism.  This means that, if we apply a diffeomorphism to the solution $\Psi$ so we have $\varphi^*\Psi$ and go through the exercise of plotting $\Gamma$ again then we get the same surface ($\Gamma$ does not move under a diffeomorphism). Hence, $\Gamma$, is a beable.

We can consider regions of operational space such as $\mathtt{A}$ and $\mathtt{B}$ shown in Fig.\ \ref{fig:AandBcomposed}.  We have, now, a space that can be used for compositional purposes.  It is an operational space if we assert that our basic observations are coincidences in the values of the set of nominated scalars.

We can define operations for General Relativity as follows.
\[ \mathsf{A}^\mathtt{a}_\mathtt{bc} = \big(\mathbf{Q}_\mathtt{A}, \mathtt{A}, \Gamma_\mathtt{A}, (\mathtt{a}, \mathtt{bc})\big)  \]
Here $\mathtt{A}$ is a region of operational space as illustrated in Fig.\ \ref{fig:AandBcomposed}, $\mathbf{Q}_\mathtt{A}$ are settings (it is beyond the scope of this paper to describe how these can be implemented in General Relativity - see \cite{hardy2016operational} for details), and $\Gamma_\mathtt{A}=\mathtt{A}\cap\Gamma$ is the outcome (it is what we see in region $\mathtt{A}$). Finally, $(\mathtt{a}, \mathtt{bc})$, are the handles.  Here the handles are given by specifying boundary surfaces for the region $\mathtt{A}$. Thus, $\mathtt{a}$ may be the surface where $\mathtt{A}$ and $\mathtt{B}$ meet.  We also equip these boundary surfaces with a (conventional) choice of direction. If the direction points away from the region associated with the operation we write it on the left of the comma and if it points towards the enclosed region we write it on the right. So, in our example $\mathtt{a}$ points away from $\mathtt{A}$ while $\mathtt{b}$ and $\mathtt{c}$ point towards $\mathtt{A}$.  We include the handles as superscripts and subscripts in our notation for the operation.  This helps with writing composite objects. For example, we may have
\[ \mathsf{A}^\mathtt{a}_\mathtt{bc} \mathsf{B}_\mathtt{a}^\mathtt{de}            \]
for the composite region shown in Fig.\ \ref{fig:AandBcomposed} whose components are joined at $\mathtt{a}$.  We can represent objects using diagrammatic notation instead. Thus we can write $\mathsf{A}^\mathtt{a}_\mathtt{bc}$ as
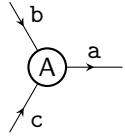
\[
\begin{Compose}{0}{0} \setdefaultfont{\mathsf}\setsecondfont{\mathtt}
\Ucircle{A}{0,0}\thispoint{UL}{120:4} \thispoint{DL}{-120:4} \thispoint{R}{0:4}
\joincc[above right]{UL}{-60}{A}{120} \csymbolalt{b}
\joincc[below right]{DL}{60}{A}{-120} \csymbolalt{c}
\joincc[above]{A}{0}{R}{180} \csymbolalt{a}
\end{Compose}
\]
Then we can write the composite object in Fig.\ \ref{fig:AandBcomposed} as
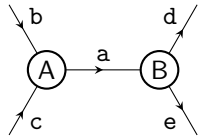
\[
\begin{Compose}{0}{0} \setdefaultfont{\mathsf}\setsecondfont{\mathtt}
\Ucircle{A}{0,0}\thispoint{UL}{120:4} \thispoint{DL}{-120:4}
\Ucircle{B}{6,0} \thispoint{UR}{$(6,0)+(60:4)$} \thispoint{DR}{$(6,0)+(-60:4)$}
\joincc[above right]{UL}{-60}{A}{120} \csymbolalt{b}
\joincc[below right]{DL}{60}{A}{-120} \csymbolalt{c}
\joincc[above]{A}{0}{B}{180} \csymbolalt{a}
\joincc[above left]{B}{60}{UR}{-120} \csymbolalt{d}
\joincc[below left]{B}{-60}{DR}{120} \csymbolalt{e}
\end{Compose}
\]
The diagrammatic notation is equivalent to the symbolic notation.

In \cite{hardy2016operational} this approach is used to provide an operational formulation of General Relativity. In the first place a possibilistic formulation is given. In this composite operations can be converted into calculations that say whether a particular configuration is possible or not.  This enables us to think about General Relativity in operational terms (kind of like thinking about the theory from a users perspective who is inside the world rather than outside it). In the second place we provide a probabilistic formulation for those cases where we have probabilistic ignorance.

\subsection{Operational space in QG}

The preceding example of operational space pertained to General Relativity. However, we can expect that the basic idea will pass over into Quantum Gravity.  In Quantum Theory we nominate some set of quantities to correspond to outcomes of our measurements.  These quantities are described by variables from classical physics (such as the position of pointers on measurement apparatuses).  Likewise, we can expect that there will be something like measurement outcomes in Quantum Gravity and that these will be described by variables from the classical theory (General Relativity in this case).  We propose that these classical quantities are coincidences in the values of scalars.  These scalars may be related to objects in the underlying theory of Quantum Gravity in a different way than they are in the classical theory.

It is possible to imagine other choices of operational space.  For example, we could relate the axes of of our operational space to quantities measured by some given set of instruments.  If these instruments have physical extension in space and time (such as measuring rods and clocks) then they will not be simply related to scalar coincidences.

An alternative approach I pursued in \cite{hardy2005probability, hardy2007towards} is to define the operational space directly from recorded data.  In that case data was imagined to be stored on cards.  A region of operational space corresponded to a subset of all possible cards.

\section{Step four: Formalism locality}

\begin{center} \fbox{\text{Formalism locality} } \end{center}
\begin{center} analogous to \end{center}
\begin{center} \fbox{ \text{Local physics}} \end{center}

\subsection{A desideratum}

We want to use a physical theory to make predictions.  Imagine we are interested in predictions pertaining to some region in our compositional space.  How are we to proceed?  In theories having a fixed causal structure we typically refer to this causal structure in doing calculations. For example, imagine in such a case we are interested in predictions pertaining to a region $\mathtt{A}$.  We can find a bigger region, $\mathtt{A}'$ (such that $\mathtt{A}\subseteq \mathtt{A}'$), having the properties that (i) $\mathtt{A}'$ has an initial spacelike hypersurface and (ii) the backward light cone of every point in $\mathtt{A}'$ is partitioned by this initial spacelike hypersurface.  In this case, we can evolve a state defined on the initial spacelike hypersurface through all of $\mathtt{A}'$. In this way, we can make predictions about $\mathtt{A}$. There are two less than satisfactory aspects to this.  First, we need to have a fixed causal structure. Second, we need to consider points outside of $\mathtt{A}$ in order to make predictions about $\mathtt{A}$.   If we do not have fixed causal structure then how are we to proceed? Then we cannot even find an appropriate region $\mathtt{A}'$ to apply this technique.  Clearly a different idea is needed.

First let us state a desideratum.  We want to formulate physical theories (whether they have definite causal structure or not) such that the formalism has the following property.
\begin{quote}
\emph{Formalism locality:}  A formulation of a physical theory is said to have the property of formalism locality if it is such that all predictions the theory can make pertaining to some given region (of the compositional space) can be made by referring only to mathematical objects (which we will call \emph{generalized states}) in the theory pertaining to this region.
\end{quote}
Clearly this dictates against using the above approach of finding a larger causally well formed region that encloses the given region.

\subsection{Prediction heralding}

At first sight it appears that there is an insurmountable obstacle to formulating a theory in a formalism local way.  We cannot use the notion of an initial state to take account of all external influences. In other words, we cannot use causal structure as an aid to screening off against influences outside the region of interest. When we have indefinite causal structure there is a sense in which regions are fundamentally open.   Consider trying to make a prediction of the general form
\[ \mathrm{Pred}_\mathtt{A}(o_\mathtt{A},o'_\mathtt{A}, s_\mathtt{A}, s'_\mathtt{A})  \]
Here $o_\mathtt{A}$ and $o'_\mathtt{A}$ are outcomes that may happen in region $\mathtt{A}$ and $s_\mathtt{A}$ and $s'_\mathtt{A}$ are some settings we choose in region $\mathtt{A}$. Our predictions could be probabilistic. In this case we might, for example, be interested in the relative probability of seeing outcome $o_\mathtt{A}$ with setting $s_\mathtt{A}$ verses seeing outcome $o'_\mathtt{A}$ with setting $s'_\mathtt{A}$ (this makes most sense when $s'_\mathtt{A}=s_\mathtt{A}$ as then we can just divide the number of times $o_\mathtt{A}$ happens by the number of times $o'_\mathtt{A}$ happens).   Or our predictions could be possibilistic. We might be interested in whether it is possible to have outcome $o_\mathtt{A}$ with setting $s_\mathtt{A}$  given that it is possible to have outcome $o'_\mathtt{A}$ with setting $s'_\mathtt{A}$.  Rather than probabilistic or possibilistic, we might have some other kind of of prediction.

The problem is that the theory may simply fail to be able to make a prediction.  This might be the case if, for example, external influences can effect what happens inside $\mathtt{A}$ since then, whatever prediction we write down, an external adversary could send in some external influence to make our prediction wrong.  When we have definite causal structure we can, under certain circumstances, guard against this. For example we can have a region such as region $\mathtt{A}'$ above (with an initial spacelike hypersurface with all other points in its domain of dependance) and we let $o'_\mathtt{A}$ and $s'_\mathtt{A}$ correspond to conditions that effectively fix the initial state.  Under these circumstances we can then make predictions about the probability or possibility of $o_\mathtt{A}$ with settings $s_\mathtt{A}$.  This is a very particular type of situation.  Generically, regions will not have this kind of special shape.  Further, if we have indefinite causal structure then we cannot, in general, even set up regions like this nor can we have an initial state.  In this case we need a different tactic to guard against meddling external adversaries.  The way forward is to give up on always being able to make predictions and then mathematise the question as to whether a prediction is possible. External adversaries will be able to mess up some predictions we might try to make but not necessarily all of them.  We need a mathematical criterion telling us when we can make predictions. We will call this \emph{prediction heralding}. If this criterion is satisfied then we need a calculation saying what the actual prediction is. This two stage process
\[ \text{Prediction Heralding} ~~~ \text{then}~~~ \text{Prediction Calculation}   \]
seems to be the only way to implement formalism locality. If we have indefinite causal structure then this way of doing local physics is pretty much forced on us.

\subsection{Causaloid implementation of prediction heralding}\label{sec:causaloid}

How can we actually implement prediction heralding?  I first discussed how to do this in the context of the causaloid framework \cite{hardy2005probability, hardy2007towards}. The same approach appears in subsequent papers setting out the duotensor framework \cite{hardy2010formalism}, the operator tensor formulation of Quantum Theory \cite{hardy2011reformulating, hardy2012operator}, and the operational formulation of General Relativity \cite{hardy2016operational}.  Let me here outline how this works in the causaloid framework. The causaloid framework is a General Probabilistic Theory (GPT) approach for indefinite causal structure.  First imagine that, for a sufficiently big region, $\mathtt{U}$, we are able to make sensible probabilistic predictions so that all probabilities of the form
\begin{equation}\label{univProb} \text{Prob}_\mathtt{U}(o_\mathtt{U}|s_\mathtt{U})   \end{equation}
have values in the theory.  This assumption can be dropped later once we have the mathematical framework in place.  Now consider a region $\mathtt{A}\subset \mathtt{U}$. We can write \eqref{univProb} as
\begin{equation}\label{probrp}
\text{Prob}_\mathtt{U}(o_\mathtt{A}, o_\mathtt{U-A}|s_\mathtt{A}, s_\mathtt{U-A})
= \mathbf{r}_\mathtt{A}[o_\mathtt{A}|s_\mathtt{A}]\cdot \mathbf{p}_\mathtt{U-A}[o_\mathtt{U-A}|s_\mathtt{U-A}]
\end{equation}
Here we can think of $\mathbf{p}_\mathtt{U-A}[o_\mathtt{U-A}|s_\mathtt{U-A}]$ as the state \lq\lq prepared" in $\mathtt{A}$ when we have outcomes $o_\mathtt{U-A}$ and settings $s_\mathtt{U-A}$ in region $\mathtt{U-A}$. The entries of the vector $\mathbf{p}_\mathtt{U-A}$ are the probabilities
\[ \text{Prob}_\mathtt{U}(o^k_\mathtt{A}, o_\mathtt{U-A}|s^k_\mathtt{A}, s_\mathtt{U-A})  \]
for a special fiducial set of \lq\lq measurements" in $\mathtt{A}$ associated with outcome setting pairs $[o^k_\mathtt{A}|s^k_\mathtt{A}]$ where $k\in\Omega_\mathtt{A}$ labels these fiducial measurements.  We put inverted commas round \lq\lq prepared" and \lq\lq measurements" since there need not be any sense in which $\mathtt{U-A}$ is before $\mathtt{A}$ - these are just words whose usefulness derives from our familiarity with them in other situations.  This fiducial set of measurements is selected so that we can write the probability in \eqref{probrp} as a dot product as shown.  We can always do this since, in the worst case, we could have a fiducial measurement for every possible $[o_\mathtt{A}|s_\mathtt{A}]$ and then the vector $\mathbf{r}_\mathtt{A}[o_\mathtt{A}|s_\mathtt{A}]$ would just have a 1 in the corresponding position and 0's everywhere else. In general, we expect the probabilities to be related to each other so that we can choose a smaller set of fiducial measurements.  We assume that we choose a minimal set (in the sense that no set of measurements of smaller rank can constitute a fiducial set).   This set up is a standard trick in the GPT approach (see \cite{hardy2001quantum} for example).  There is a duality between $\mathbf{r}$ and $\mathbf{p}$ vectors. We could equally well have set this up with a $\mathbf{p}_\mathtt{A}[o_\mathtt{A}|s_\mathtt{A}]$ for $\mathtt{A}$ and an $\mathbf{r}_\mathtt{U-A}[o_\mathtt{U-A}|s_\mathtt{U-A}]$ for $\mathtt{U-A}$.  These objects (whether in the $\mathbf{p}$ or $\mathbf{r}$ form) are \emph{generalized states}.

Now we can see how to implement formalism locality with a two stage calculation.  Let the thing we want to calculate be the relative probability
\begin{equation}\label{predprob}
 \mathrm{Pred}_\mathtt{A}(o_\mathtt{A},o'_\mathtt{A}, s_\mathtt{A}, s'_\mathtt{A})
 = \frac{\mathrm{Prob}(o_\mathtt{A}|s_\mathtt{A})}{\mathrm{Prob}(o'_\mathtt{A}|s'_\mathtt{A})}
\end{equation}
pertaining to region $\mathtt{A}$.  Something we can calculate  (pertaining to region $\mathtt{U}$) is
\[
\frac{\mathrm{Prob}(o_\mathtt{A}, o_\mathtt{U-A}|s_\mathtt{A},s_\mathtt{U-A} )}{\mathrm{Prob}(o'_\mathtt{A},o_\mathtt{U-A}|s'_\mathtt{A},s_\mathtt{U-A})}
=
\frac{
\mathbf{r}_\mathtt{A}[o_\mathtt{A}|s_\mathtt{A}]\cdot \mathbf{p}_\mathtt{U-A}[o_\mathtt{U-A}|s_\mathtt{U-A}]
}{
\mathbf{r}_\mathtt{A}[o'_\mathtt{A}|s'_\mathtt{A}]\cdot \mathbf{p}_\mathtt{U-A}[o_\mathtt{U-A}|s_\mathtt{U-A}]
}
\]
We have the extra arguments $o_\mathtt{U-A}$ and $s_\mathtt{U-A}$ pertaining to region $\mathtt{U-A}$ so this prediction is not of the form in \eqref{predprob}.  If we are to have formalism locality our predictions must be independent of $o_\mathtt{U-A}$ and $s_\mathtt{U-A}$.  This means the prediction must be independent of $\mathbf{p}_\mathtt{U-A}[o_\mathtt{U-A}|s_\mathtt{U-A}]$.  This will be true if and only if the following condition holds
\begin{equation}\label{predherald}
\mathbf{r}_\mathtt{A}[o_\mathtt{A}|s_\mathtt{A}] \propto \mathbf{r}_\mathtt{A}[o'_\mathtt{A}|s'_\mathtt{A}]      ~~~~~~ \text{Prediction Heralding}
\end{equation}
This is the prediction heralding condition.  In the case the condition does hold then
\begin{equation}
\frac{\mathrm{Prob}(o_\mathtt{A}|s_\mathtt{A})}{\mathrm{Prob}(o'_\mathtt{A}|s'_\mathtt{A})} = k ~~~~~~ \text{Prediction Calculation}
\end{equation}
where $k$ is given by
\begin{equation}
\mathbf{r}_\mathtt{A}[o_\mathtt{A}|s_\mathtt{A}] = k\mathbf{r}_\mathtt{A}[o'_\mathtt{A}|s'_\mathtt{A}]
\end{equation}
This shows how to implement formalism locality in probabilistic theories.  In possibilistic theories a similar strategy can be pursued (see \cite{hardy2016operational} for an example).

\subsection{Replace trade-craft with calculation}

In standard formulations of physical theories there is a certain trade-craft to knowing what quantities we can make predictions for. As physicists, we generally have a good intuition for this without fully appreciating that we are only considering a tiny subset of possible predictions for which questions can be posed.  This trade-craft invariably involves looking at the causal structure operating in the background. The idea here is to liberate ourselves from this intuition led approach and consider predictions of any sort. Then we have a mathematical criterion for prediction heralding.  This leads to formalism locality and the two stage calculation process and, indeed, this is pretty much forced on us by indefinite causal structure. However, even when we have definite causal structure it may be argued that the approach argued for here is superior. In particular, this approach is democratic with regard to what types of predictions it will attempt to make.

\subsection{A fallacy}

One fallacy regarding the formalism local approach here is that it is much worse than the \lq\lq standard" approach as predictions are only heralded in the very non-generic situation in which two vectors are parallel.  To respond to this let me first point out that any situation in the standard approach to physics where we can make a prediction will, when translated into these terms, necessarily have these two vectors parallel (since this was a necessary condition for the heralding of a prediction).  Hence the non-generic issue is just as much of a problem for the standard approach.  But, secondly, within the approach advocated here we can relax the prediction heralding condition (this is not open to us in the standard approach).  For example, if the two vectors are not quite parallel then, while we cannot give an exact prediction for the relative probability in \eqref{predprob}, we can put some bounds on this prediction.  For an outline of how to do this see \cite{markes2009entropy}.

\section{Step five: Laws given by correspondence map}\label{sec:lawsgivenbycorr}

\begin{center}  \fbox{\text{Laws given by correspondence map} }  \end{center}
\begin{center}  analogous to  \end{center}
\begin{center} \fbox{\text{Laws expressed by field equations}}   \end{center}

\noindent
The way a physical law is expressed depends on the framework within which the theory is given.  In General Relativity we provide the laws of physics by giving a set of field equations that constrain the fields.  In operational formulations the laws are given by providing a map from operations (that describe the operational aspects of the real world) to generalized states (that are used in calculations to make predictions)
\[ \text{operation} \rightarrow \text{generalized state}   \]
If we denote the operation associated with region $\mathtt{A}$ by $\mathsf{A}$ and the generalized state by $A$  then the correspondence map is
\[ \mathsf{A} \rightarrow A \]
To specify the laws we need to provide the equations that give this map.

The next principle will concern composite regions and how to combine generalized states for them.  Thus, it is sufficient to express the laws (by giving correspondence maps) for small regions since then we can obtain generalized states for larger regions. We could choose some set of small (but finite) regions for this purpose.  Or we might choose to take the limit as these regions become infinitesimal.  In the latter case we only require a correspondence map for infinitesimally small regions.  Schematically, we can represent this for each point, $\mathbf{S}$, in compositional space
\[ \mathsf{A}(\mathbf{S}, \delta\mathbf{S}) \overset{f_\mathbf{S}}{\rightarrow}  \hat{A} (\mathbf{S}, \delta\mathbf{S})  \]
where $\delta\mathbf{S}$ generates a small volume about $\mathbf{S}$.  Now the law is given by $f_\mathbf{S}$.  How to implement this infinitesimal schema will be the subject of further research.

\section{Step six: Boundary mediated compositional description}\label{sec:boundarymediated}

\begin{center} \fbox{\text{Boundary mediated compositional description}}  \end{center}
\begin{center} analogous to  \end{center}
\begin{center}  \fbox{\text{Local tensor fields based on tangent space}} \end{center}

\noindent
How do we describe a composite object?  In basic terms we should provide a list of the components and the relationships between these components.   A relationship can exist between any subset of the components.  For example, in a composite object consisting of component objects $\mathsf{A}$, $\mathsf{B}$, $\mathsf{C}$, and $\mathsf{D}$, there could in principle be relationships between all pairs, relationships between all triples, and a relationship between all four components.  The description of this object would be represented by a hypergraph having all possible hyperedges.  It could be the case, however, that some relationships are implied by other relationships. In particular, consider an object whose component objects correspond to a number of regions that border one another.  We can represent this by a graph in which edges correspond to common boundaries.  For example
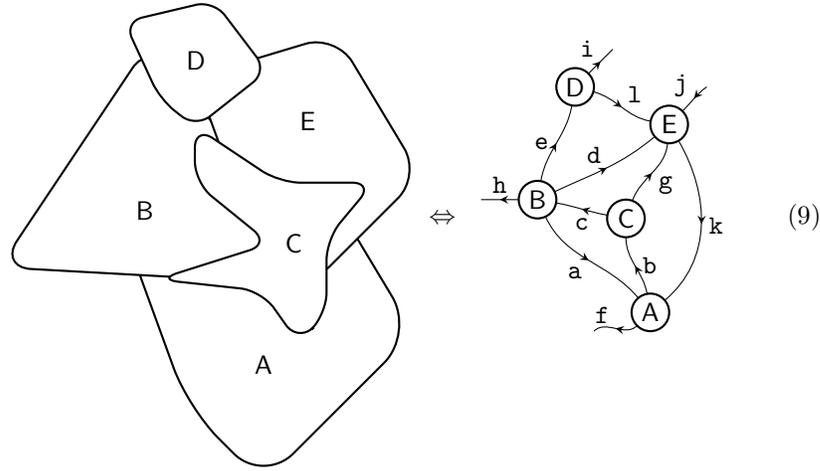
\begin{equation}\label{blobsetc}
\begin{Compose}{0}{0}
\cobjectwhite{\theblob}{A}{2}{2}{-0.2,-5} \csymbol[0,-90]{A}
\cobjectwhite{\theflag}{B}{2}{1}{-4.75,0} \csymbol[-50,0]{B}
\cobjectwhite{\theblob}{E}{1.5}{1.5}{2.15,3} \csymbol[0,50]{E}
\cobjectwhite{\thetooth}{C}{1}{1}{1.4,-1.75} \csymbol{C}
\cobjectwhite{\theblob}{D}{0.95}{0.75}{-3.8,8} \csymbol{D}
\end{Compose}
\Leftrightarrow ~~
\begin{Compose}{0}{-1.1} \setdefaultfont{\mathsf}\setsecondfont{\mathtt}
\Ucircle{A}{1,-1} \Ucircle{B}{-5,5} \Ucircle{C}{-0.3,4} \Ucircle{D}{-3, 11} \Ucircle{E}{2,9}
\joincc[below left]{B}{-65}{A}{130} \csymbolalt{a}
\joincc[right]{A}{100}{C}{-90} \csymbolalt{b}
\joincc[below]{C}{170}{B}{-10} \csymbolalt{c}
\joincc[above left]{B}{25}{E}{-140} \csymbolalt{d}
\joincc[left]{B}{80}{D}{-100} \csymbolalt{e}
\joincc[above right]{D}{-15}{E}{170} \csymbolalt{l}
\joincc[below right]{C}{70}{E}{-95} \csymbolalt{g}
\joincc[right]{E}{-60}{A}{40}\csymbolalt{k}
\thispoint{nA}{-2,-2} \joincc[above left]{A}{-135}{nA}{45} \csymbolalt{f}
\thispoint{nB}{-8,5} \joincc[above]{B}{180}{nB}{0} \csymbolalt{h}
\thispoint{nD}{-1,13} \joincc[above left]{D}{45}{nD}{-135} \csymbolalt{i}
\thispoint{nE}{4,11} \joincc[above left]{nE}{-145}{E}{45} \csymbolalt{j}
\end{Compose}
\end{equation}
In the graph on the left we only specify relationships between regions that border one another.  We also include open boundaries (such as $\mathtt{h}$ in the above example) since we may consider joining this composite object to other objects.  The arrows indicate that we can associate a conventional choice of direction with the boundary surfaces (they do not indicate the flow of time).  We can use symbolic (rather than diagrammatic) notation to represent this graph
\[
\mathtt{
\mathsf{A}_{ak}^{fb} \mathsf{B}_{c}^{heda}
\mathsf{C}_{b}^{cg} \mathsf{D}_{e}^{il} \mathsf{E}_{dgjl}^{k}
}
\]
In this case, subscripts correspond to boundaries with the direction pointing in and superscripts to boundaries with the direction pointing out.

We will call this \emph{boundary mediated compositional description}.  The point is that we adopt this form of description for composite objects.  This matters when we adopt the principle of general compositionality.

\section{Step seven: The principle of general compositionality}

\begin{center}  \fbox{\text{Principle of general compositionality} }  \end{center}
\begin{center}  analogous to  \end{center}
\begin{center} \fbox{\text{Principle of general covariance}}   \end{center}

\subsection{The principle}

We can break a big region of compositional space into smaller component regions in many different ways. In the absence of definite causal structure, no particular way of breaking up compositional space is preferred. Objects are associated with regions of compositional space. Consequently, we need to write the laws of physics so that they work for any decomposition of an object into component objects.  Furthermore, in the absence of definite causal structure, the structure of these calculations can only be informed by the compositional description of these objects. Hence, it is reasonable to assume that the calculation has the same compositional structure.  These considerations lead to the following:
\begin{quote}
\textbf{The principle of general compositionality}: The laws of physics should be written in such a way that they work for any decomposition of an object into components
and so that the calculations for any such decomposition have the same compositional structure as the compositional description of the object for this decomposition.
\end{quote}
It is understood that the compositional description used here is the boundary mediated compositional description outlined in Sec.\ \ref{sec:boundarymediated}.
This principle is deliberately written in similar language to Einstein's principle of general covariance: the laws of physics should be written in such a way that they work to any coordinate system and so that they take the same form in any coordinate system.  The motivation for the principle of general compositionality (from the absence of definite causal structure) is strongly analogous to the motivation for Einstein's principle (from the absence of a global inertial reference frame (as outlined in Sec.\ \ref{sec:sevensteps})).

\subsection{Correspondence}

In Sec.\ \ref{sec:lawsgivenbycorr} we saw how laws are given by providing a correspondence map from operations to generalized states.  In Sec.\ \ref{sec:boundarymediated} we adopted a boundary mediated compositional description.  To formulate the laws of physics in accord with the principle of general compositionality we need, then, to construe our generalized states in such a way that they are related to the boundary mediated descriptions of objects. We need a \emph{correspondence principle} of the form
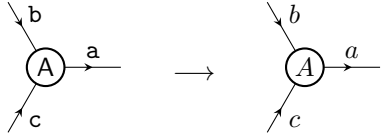
\[
\begin{Compose}{0}{0} \setdefaultfont{\mathsf}\setsecondfont{\mathtt}
\Ucircle{A}{0,0}\thispoint{UL}{120:4} \thispoint{DL}{-120:4} \thispoint{R}{0:4}
\joincc[above right]{UL}{-60}{A}{120} \csymbolalt{b}
\joincc[below right]{DL}{60}{A}{-120} \csymbolalt{c}
\joincc[above]{A}{0}{R}{180} \csymbolalt{a}
\end{Compose}
~~~~~\longrightarrow~~~~~
\begin{Compose}{0}{0} \setdefaultfont{\mathnormal}\setsecondfont{\mathnormal}
\Ucircle{A}{0,0}\thispoint{UL}{120:4} \thispoint{DL}{-120:4} \thispoint{R}{0:4}
\joincc[above right]{UL}{-60}{A}{120} \csymbolalt{b}
\joincc[below right]{DL}{60}{A}{-120} \csymbolalt{c}
\joincc[above]{A}{0}{R}{180} \csymbolalt{a}
\end{Compose}
\]
from objects to generalized states.  This takes us beyond the causaloid formalism discussed in Sec.\ \ref{sec:causaloid}.  We will discuss two ways of so representing generalized states.  First, by a \emph{duotensor} \cite{hardy2010formalism} and second by an \emph{operator tensor} \cite{hardy2011reformulating}. Both the duotensor and operator tensor are linear in $\mathbf{r}_\mathtt{A}$ discussed in Sec.\ \ref{sec:causaloid}.  However, crucially, they are expressed with respect to boundary information - they provide a way to understand composition that is missing in the causaloid formalism.

For general probability theories having the property of tomographic locality the generalized state can be represented by a \emph{duotensor}
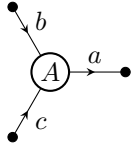
\[
\begin{Compose}{0}{0} \setdefaultfont{\mathnormal}\setsecondfont{\mathnormal}
\Ucircle{A}{0,0}\blackdot{UL}{120:4} \blackdot{DL}{-120:4} \blackdot{R}{0:4}
\joincc[above right]{UL}{-60}{A}{120} \csymbolalt{b}
\joincc[below right]{DL}{60}{A}{-120} \csymbolalt{c}
\joincc[above]{A}{0}{R}{180} \csymbolalt{a}
\end{Compose}
\]
Something similar can be done for general possibilistic theories (having a similar tomographic locality property) \cite{hardy2016operational}.  In possibilistic theories we calculate whether something is possible (which we give value 1) or impossible (which we give value 0).  A duotensor is like a tensor but with a bit more structure.  We can have two positions for each superscript and each subscript in symbolic notation. In diagrammatic notation we represent these two positions by using black or white dots.  Physically, the duotensor with all black dots corresponds to the probability when we putting fiducial objects (labeled by $a=1$ to $K_\mathtt{a}$) on each boundary.  In the case of Quantum Theory we can use these duotensors to linearly weight fiducial operators to construct \emph{operator tensors} having the form
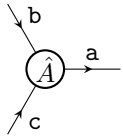
\[
\begin{Compose}{0}{0} \setdefaultfont{\hat}\setsecondfont{\mathtt}
\Ucircle{A}{0,0}\thispoint{UL}{120:4} \thispoint{DL}{-120:4} \thispoint{R}{0:4}
\joincc[above right]{UL}{-60}{A}{120} \csymbolalt{b}
\joincc[below right]{DL}{60}{A}{-120} \csymbolalt{c}
\joincc[above]{A}{0}{R}{180} \csymbolalt{a}
\end{Compose}
\]
As discussed in Sec.\ \ref{sec:optensQT}, an operator tensor is an Hermitean operator acting on a Hilbert space determined by the labels on the legs ($\mathcal{H}^\mathtt{a}\otimes\mathcal{H}_\mathtt{b}\otimes\mathcal{H}_\mathtt{c}$ in this case).

\subsection{Implementation}

Once we have generalized states of the above form we can do a calculation.  We have already learned that,  in order to make predictions for some region, we need the relevant generalized states. So we need to know how to calculate the generalized state for a composite operation.  Let us consider the example in \eqref{blobsetc}.  In the duotensor formalism the generalized state associated with this composite operation is given by replacing each operation with the corresponding duotensor:
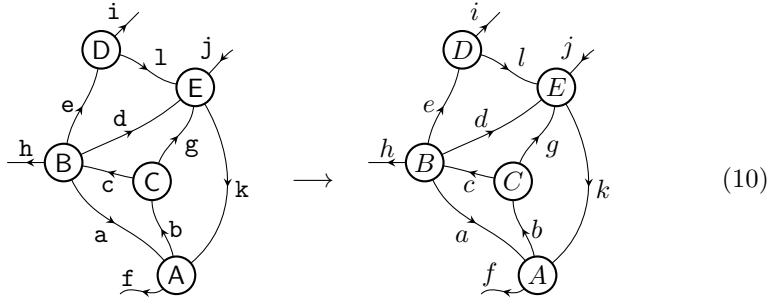
\begin{equation}\label{duotensorcalc}
\begin{Compose}{0}{-1.1} \setdefaultfont{\mathsf}\setsecondfont{\mathtt}
\Ucircle{A}{1,-1} \Ucircle{B}{-5,5} \Ucircle{C}{-0.3,4} \Ucircle{D}{-3, 11} \Ucircle{E}{2,9}
\joincc[below left]{B}{-65}{A}{130} \csymbolalt{a}
\joincc[right]{A}{100}{C}{-90} \csymbolalt{b}
\joincc[below]{C}{170}{B}{-10} \csymbolalt{c}
\joincc[above left]{B}{25}{E}{-140} \csymbolalt{d}
\joincc[left]{B}{80}{D}{-100} \csymbolalt{e}
\joincc[above right]{D}{-15}{E}{170}\csymbolalt{l}
\joincc[below right]{C}{70}{E}{-95} \csymbolalt{g}
\joincc[right]{E}{-60}{A}{40}\csymbolalt{k}
\thispoint{nA}{-2,-2} \joincc[above left]{A}{-135}{nA}{45} \csymbolalt{f}
\thispoint{nB}{-8,5} \joincc[above]{B}{180}{nB}{0} \csymbolalt{h}
\thispoint{nD}{-1,13} \joincc[above left]{D}{45}{nD}{-135} \csymbolalt{i}
\thispoint{nE}{4,11} \joincc[above left]{nE}{-145}{E}{45} \csymbolalt{j}
\end{Compose}
~~~\longrightarrow~~~
\begin{Compose}{0}{-1.1} \setdefaultfont{\mathnormal}\setsecondfont{\mathnormal}
\Ucircle{A}{1,-1} \Ucircle{B}{-5,5} \Ucircle{C}{-0.3,4} \Ucircle{D}{-3, 11} \Ucircle{E}{2,9}
\joincc[below left]{B}{-65}{A}{130} \csymbolalt{a}
\joincc[right]{A}{100}{C}{-90} \csymbolalt{b}
\joincc[below]{C}{170}{B}{-10} \csymbolalt{c}
\joincc[above left]{B}{25}{E}{-140} \csymbolalt{d}
\joincc[left]{B}{80}{D}{-100} \csymbolalt{e}
\joincc[above right]{D}{-15}{E}{170} \csymbolalt{l}
\joincc[below right]{C}{70}{E}{-95} \csymbolalt{g}
\joincc[right]{E}{-60}{A}{40}\csymbolalt{k}
\thispoint{nA}{-2,-2} \joincc[above left]{A}{-135}{nA}{45} \csymbolalt[0,6]{f}
\thispoint{nB}{-8,5} \joincc[above]{B}{180}{nB}{0} \csymbolalt{h}
\thispoint{nD}{-1,13} \joincc[above left]{D}{45}{nD}{-135} \csymbolalt{i}
\thispoint{nE}{4,11} \joincc[above left]{nE}{-145}{E}{45} \csymbolalt{j}
\end{Compose}
\end{equation}
To do the calculation shown on the left we sum over indices where there is a wire drawn between duotensors (this is essentially Penrose diagrammatic notation).
Once we have this generalized state, we can use it to do calculations (prediction heralding and prediction calculation) as outlined in Sec.\ \ref{sec:causaloid} (since it is linearly related to $\mathbf{r}_\mathtt{A}$, the details work the same way).   We see that the calculation has the same structure as the compositional description. Furthermore, we can calculate this generalized state with respect to any decomposition into components. Hence the principle of generalized compositionality is satisfied.

If we represent generalized states by operator tensors then we have
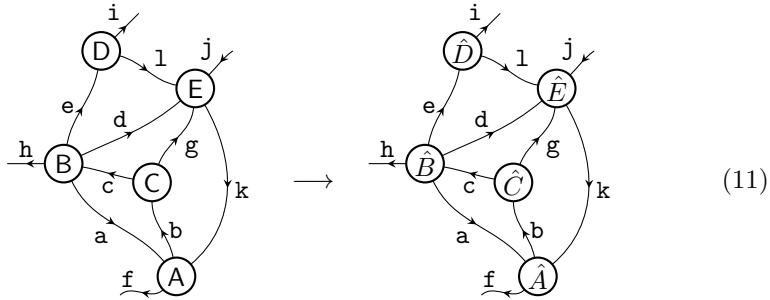
\begin{equation}\label{operatortensorcalc}
\begin{Compose}{0}{-1.1} \setdefaultfont{\mathsf}\setsecondfont{\mathtt}
\Ucircle{A}{1,-1} \Ucircle{B}{-5,5} \Ucircle{C}{-0.3,4} \Ucircle{D}{-3, 11} \Ucircle{E}{2,9}
\joincc[below left]{B}{-65}{A}{130} \csymbolalt{a}
\joincc[right]{A}{100}{C}{-90} \csymbolalt{b}
\joincc[below]{C}{170}{B}{-10} \csymbolalt{c}
\joincc[above left]{B}{25}{E}{-140} \csymbolalt{d}
\joincc[left]{B}{80}{D}{-100} \csymbolalt{e}
\joincc[above right]{D}{-15}{E}{170} \csymbolalt{l}
\joincc[below right]{C}{70}{E}{-95} \csymbolalt{g}
\joincc[right]{E}{-60}{A}{40}\csymbolalt{k}
\thispoint{nA}{-2,-2} \joincc[above left]{A}{-135}{nA}{45} \csymbolalt{f}
\thispoint{nB}{-8,5} \joincc[above]{B}{180}{nB}{0} \csymbolalt{h}
\thispoint{nD}{-1,13} \joincc[above left]{D}{45}{nD}{-135} \csymbolalt{i}
\thispoint{nE}{4,11} \joincc[above left]{nE}{-145}{E}{45} \csymbolalt{j}
\end{Compose}
~~~\longrightarrow~~~
\begin{Compose}{0}{-1.1} \setdefaultfont{\hat}\setsecondfont{\mathtt}
\Ucircle{A}{1,-1} \Ucircle{B}{-5,5} \Ucircle{C}{-0.3,4} \Ucircle{D}{-3, 11} \Ucircle{E}{2,9}
\joincc[below left]{B}{-65}{A}{130} \csymbolalt{a}
\joincc[right]{A}{100}{C}{-90} \csymbolalt{b}
\joincc[below]{C}{170}{B}{-10} \csymbolalt{c}
\joincc[above left]{B}{25}{E}{-140} \csymbolalt{d}
\joincc[left]{B}{80}{D}{-100} \csymbolalt{e}
\joincc[above right]{D}{-15}{E}{170} \csymbolalt{l}
\joincc[below right]{C}{70}{E}{-95} \csymbolalt{g}
\joincc[right]{E}{-60}{A}{40}\csymbolalt{k}
\thispoint{nA}{-2,-2} \joincc[above left]{A}{-135}{nA}{45} \csymbolalt{f}
\thispoint{nB}{-8,5} \joincc[above]{B}{180}{nB}{0} \csymbolalt{h}
\thispoint{nD}{-1,13} \joincc[above left]{D}{45}{nD}{-135} \csymbolalt{i}
\thispoint{nE}{4,11} \joincc[above left]{nE}{-145}{E}{45} \csymbolalt{j}
\end{Compose}
\end{equation}
The object on the right is the generalized state, expressed as an operator tensor, for the composite region.  We saw in Sec.\ \ref{sec:optensQFT} how to formulate Quantum Field Theory in this way.  That formulation assumed a fixed background manifold with a fixed metric (and time direction field).  What we do not have yet is a formulation of Quantum Gravity in terms of operator tensors.  To do this we would have to go other to using compositional space (rather than manifold space) and identify the appropriate fiducial objects to formulate the theory. The operational formulation of classical General Relativity given in \cite{hardy2016operational} is, likely, a good stepping stone for this purpose.

\subsection{Remarks}

In 1917 Kretchman criticized Einstein's principle of general covariance on the grounds that it is possible to formulate any physical theory in such terms \cite{kretschmann1917uber}. Einstein's response \cite{Einstein1918prinzipielles} was that, whilst this may be true, the principle plays a powerful heuristic role in obtaining the theory of General Relativity and that one should prefer the simpler and more transparent case.  Similarly remarks may apply for the principle of general compositionality.  It is possible to formulate quantum theory, classical probability theory, and possibly any theory in accord with this principle. However, we may hope that the correct theory of Quantum Gravity emerges as simplest and most elegant such theory.

In setting up the principle of general compositionality we adopted a particular way of describing composite objects - namely the boundary based description outlined in Sec.\ \ref{sec:boundarymediated}.  However, as was pointed out, there are many other ways of describing composite objects which may be equally valid.  This raises the possibility of a more general principle.
The general principle of general compositionality would say: The laws of physics should be written in such a way that they work for any decomposition of an object up into components and so that the calculations for any such decomposition can be written so that they have the same compositional structure as any way of providing the compositional description of the object for this decomposition. It is not clear whether this more general principle can be implemented or whether there is anything to be gained in doing so.

\section{Three possible elements of Quantum Gravity}

The first seven steps provide a possible framework for Quantum Gravity. The basic shape of this framework is established though there remains more work to be done.  The real task, however, is to provide analogues for the three elements of General Relativity that will define the theory of Quantum Gravity.   At this stage I have only basic ideas.

\subsection*{First element: A prescription}\label{sec:firstelement}

If we start with Quantum Field theory then we will have a correspondence map from objects to generalized states (operator tensors).
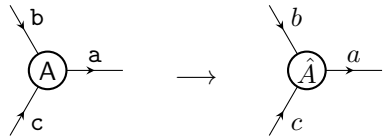
\[
\begin{Compose}{0}{0} \setdefaultfont{\mathsf}\setsecondfont{\mathtt}
\Ucircle{A}{0,0}\thispoint{UL}{120:4} \thispoint{DL}{-120:4} \thispoint{R}{0:4}
\joincc[above right]{UL}{-60}{A}{120} \csymbolalt{b}
\joincc[below right]{DL}{60}{A}{-120} \csymbolalt{c}
\joincc[above]{A}{0}{R}{180} \csymbolalt{a}
\end{Compose}
~~~~~\longrightarrow~~~~~
\begin{Compose}{0}{0} \setdefaultfont{\hat}\setsecondfont{\mathnormal}
\Ucircle{A}{0,0}\thispoint{UL}{120:4} \thispoint{DL}{-120:4} \thispoint{R}{0:4}
\joincc[above right]{UL}{-60}{A}{120} \csymbolalt{b}
\joincc[below right]{DL}{60}{A}{-120} \csymbolalt{c}
\joincc[above]{A}{0}{R}{180} \csymbolalt{a}
\end{Compose}
\]
Since these correspondences are maps, they are defined by some equations.  We need a prescription for turning correspondences like this coming from Quantum Field Theory into correspondences for Quantum Gravity. In so doing some gravitational degrees of freedom (analogous to the metric) will be introduced.

\subsection*{Second element: Gravitational addendum}

Since we have introduced variable gravitational degrees of freedom, we need some extra equations to fix them so we still have a correspondence from objects to generalized states.  This is the gravitational addendum.

In Quantum Field Theory, the metric is constant.  The role of the metric in the operator tensor formulation of Quantum Field Theory is to allow us to impose physicality conditions.  As discussed in Sec.\ \ref{sec:optensQFT}, these are conditions that guarantee that (i) probabilities are bounded between 0 and 1 and (ii) agents cannot signal backward in time.

In General Relativity we have a gravitational addendum in the form of Einstein's field equations.  The metric is nine tenths causality.  Hence, it seems that the analogue to Einstein's field equations in Quantum Gravity (in the approach I am proposing) is a further generalization of the physicality conditions that allows causal structure to be variable and indefinite.  In this approach to Quantum Gravity the holy grail is, therefore, to find the physicality conditions when we have such variable and indefinite causal structure.


\subsection*{Third element: An interpretation}

Since its inception people have struggled to find the correct interpretation of Quantum Theory.  Contrast this with General Relativity where there is a generally agreed upon interpretation (outlined above) that naturally emerges from the formalism.  My hope is that, once we have our hands on the correct theory of Quantum Gravity, an interpretation will naturally emerge from the formalism.  This may seem like a tall order. However, there are some grounds to be optimistic. Many of the interpretational difficulties of Quantum Theory can be traced back to the conservative role that causal structure plays in the theory (Bell's theorem, unitary evolution with respect to a fixed background time, and so on).  In a theory of Quantum Gravity the radical nature of causal structure witnessed in General Relativity will necessarily be present and may offer a way to resolve some of these interpretational difficulties.


\part{The Quantum Equivalence Principle} \label{part:thequantumep}

\section{Introduction}

The conceptual path to Quantum Gravity proposed in Part \ref{part:roadtoQG} was analogous to the path taken by Einstein in constructing General Relativity. We can, however, pursue an even closer analogy.  Here I propose the quantum equivalence principle that would play a strictly analogous role to the equivalence principle.  This path could share many elements in common with the path sketched out in Part \ref{part:roadtoQG}.  This path is, however, much more tentative as I do not have an implementation of the basic idea of a quantum frame of reference as suitable to this particular approach.

\section{Quantum Frames of Reference}

Quantum frames of reference been much studied (see, for example, \cite{aharonov1967charge, aharonov1984quantum, bartlett2007reference, pienaar2015relativity, giacomini2017quantum}). The work of Giacomini et.\ al.\  \cite{giacomini2017quantum} is particularly promising.  A related notion is a \emph{causal reference frame} suggested by Gu\'erin and Brukner \cite{guerin2018observer}.  In the causal reference frame of a particular event this event is well localized in time and other events are smeared out. Though there also exists a causal reference frame in which any of these events are localized in time.  Oreshkov's discussion of time-delocalized quantum subsystems \cite{oreshkov2018whereabouts} also bears on this issue (in particular, he also shows how to represent the circuit so that one of the parties is time-localized).   Hoehn has also explored how quantum reference frames may have a bearing on quantum gravity \cite{hoehn2015quantum}.

I will lump quantum reference frames and causal reference frames in together as it seems they respectively address the quantum and the causal notions that will be required. I will simply call the resulting concept \emph{quantum reference frames} (an alternative name might be quantum causal reference frames).  We might also refer to \emph{quantum coordinate systems} which can be used to chart \emph{quantum manifolds}.  Gu\'erin and Brukner speculate about such coordinates and whether the usual diffeomorphism symmetry group of General Relativity might be enlarged to take them into account (see comments at the end of Sec.\ 5 in \cite{guerin2018observer}). 

The million dollar question is: how should we actually implement quantum reference frames as suitable for the approach to quantum gravity I advocate here?  I do not have a good answer to this question.  However, a few points we may consider are the following.  First, we need to go beyond the idea of a quantum reference frame as pertaining to a given time in the direction of a frame that concerns different times as do causal reference frames.  Second, we want to consider the continuous case.  Third, we want to have transformations between these reference frames.  It would be useful if these transformations were smooth and invertible as are diffeomorphisms in General Relativity.  Thus, we want to have quantum diffeomorphisms.  It is worth noting, in this context, that there are ways to have smooth transformations inside Hilbert space not available in regular space. For example, we can transform smoothly from $|s_A\rangle$ to $|s_B\rangle$ when the points $s_A$ and $s_B$ are non-adjacent without going through any points in between these points simply by writing
$\cos(\theta)|s_A\rangle+\sin(\theta)|s_B\rangle$ and letting $\theta$ range from $0$ to $\pi$.  This may make quantum diffeomorphisms much more interesting (and counterintuitive) than the quantum case.  We will denote points in the quantum reference frame by $\mathbf{s}$.  In a particular realization, there may be a map from $\mathbf{s}$ to points, $\mathbf{S}$ in operational space as discussed in Sec.\ \ref{sec:compspace}.

\section{The Quantum Equivalence Principle} \label{sec:quantumequivprinc}

The (classical) equivalence principle in its strong form states that the laws of physics are the same locally in any falling frame of reference. We can state this as saying that it is always possible to transform to a reference frame in which inertial laws of physics apply in the local vicinity of any point.  This principle is motivated by Galileo's famous experiment dropping two different masses from the Leaning Tower of Pisa (though it is not known whether he actually carried out this experiment).

We state
\begin{quote}
\textbf{The quantum equivalence principle}: It is always possible to transform to a quantum reference frame in which we have definite causal structure in the local vicinity of any point.
\end{quote}
According to this principle we can transform away indefinite causal structure locally but not necessarily globally.  This is clearly analogous to the equivalence principle where we can transform into an inertial reference frame locally but not necessarily globally.

At this stage we do not have a proof that such a principle is possible. However, the above cited work provides strong evidence that this may be possible. In particular, the work of Gu\'erin and Brukner \cite{guerin2018observer} provides a kind of discrete realization (at a point but not in the vicinity of the point) with the corresponding language of causal reference frames - this is the best quantum analogue of Galileo's Leaning Tower of Pisa experiment we currently have.

There has been discussion of a quantum equivalence principle in the literature before \cite{castagnino1980quantum, candelas1983there, osipova1986quantum, hessling1994quantum, kleinert1997quantum} but those proposals have been of a very different nature to that suggested here.

\section{Local correspondence}\label{sec:localcorrespondence}

In Sec.\ \ref{sec:lawsgivenbycorr} we discussed the idea of shrinking operations and the operators corresponding to them so that they become infinitesimal wherein the correspondence map that represents the laws of physics was schematically represented as
\[ \mathsf{A}(\mathbf{s}, \delta\mathbf{s}) \overset{f_\mathbf{s}}{\rightarrow}  \hat{A} (\mathbf{s}, \delta\mathbf{s})  \]
(though now we are employing points, $\mathbf{s}$ in the quantum reference frame).
Now the physics is given by the functions $f_\mathbf{S}$ from these infinitesimal operations to the corresponding operators.  It is worth filling this out a little for our present purposes.  The function $f_\mathbf{s}$ takes the specification of the infinitesimal operation $\mathsf{A}(\mathbf{s}, \delta\mathbf{s})$ as its argument. Since this is infinitesimal, it is given by the values of the setting and outcome fields at $\mathbf{s}$ and some finite number of derivatives.  An analogous situation from quantum theory is when we determine the infinitesimal unitary evolution operator from $t$ to $t+\delta t$ as $\hat{U}(t, \delta t) = \exp(-i\hat{H}(\theta)\delta t/\hbar)$. Here the Hamiltonian is determined by the external settings, $\theta$.  In general, we might imagine there is some function analogous to $\hat{H}(\theta)$ from which we can determine $f_\mathbf{s}$.  Let us call this function
\[     F_\mathbf{s}(\text{settings, outcomes, and some derivatives})   \]
We expect that $F_\mathbf{s}$ will map its arguments to some number of real fields (since this is true of the operator, $\hat{A} (\mathbf{s}, \delta\mathbf{s})$).  Let this number be $N$.  Thus, we need $N$ equations to specify these fields in terms of the settings, outcomes and some of their derivatives.  These equations are analogous to the field equations in classical General Relativity.

\section{Seven Steps}

I propose seven steps as follows
\[
\begin{array}{ll}
{}          & \fbox{ \text{The quantum equivalence principle} } \\
\rightarrow & \fbox{\text{No global reference frame with definite causal structure} }\\
\rightarrow &\fbox{\text{General quantum coordinates}} \\
\rightarrow &\fbox{\text{Local physics in terms of fields} } \\
\rightarrow &\fbox{\text{Laws given by local correspondence functions}} \\
\rightarrow &\fbox{\text{Appropriate simple choice of mathematical objects}} \\
\rightarrow &\fbox{\text{Principle of general quantum covariance} }
\end{array}
\]
Filling in a few more details we would proceed as follows. First we establish that there exists a quantum equivalence principle as outlined above. Then we see this means we have no global reference frame with definite causal structure.  This means we need to use general quantum coordinates which we will call $\mathbf{s}$. With respect to these coordinates we write down physics in terms of the locally specified fields, $F_\mathbf{s}$ (discussed in Sec.\ \ref{sec:localcorrespondence}).  We need to make some appropriate choice of how to express these local fields.  It is possible that we can think of them as operators on local tangent spaces.  Then we demand that, $F_\mathbf{s}$, can be written in a form that is generally covariant under changes of quantum reference frame.

\section{Three elements}

If we can make the quantum equivalence principle work then we are in very good position to implement analogues to the three elements of general relativity.  The proposal is that these would be as follows.
\[
\begin{array}{ll}
\rightarrow &\fbox{\text{Prescription for turning QFT calculations into QG calculations}} \\
\rightarrow &\fbox{\text{Addendum: New physicality conditions for Quantum Gravity}} \\
\rightarrow &\fbox{\text{An interpretation}}
\end{array}
\]
We will discuss each of these in turn.

\subsection{First element: prescription}

If we can transform locally into a definite reference frame then we have QFT locally in that reference frame.  We need to find a way to modify QFT calculations such that we can get QFT locally by transforming into such a reference frame.
In the analogous case of General Relativity this consists of replacing partial derivatives by covariant derivatives, the Minkowski metric by the general metric, and replacing all remaining inertial coordinates by general coordinates.  This is called minimal substitution.   Quantum Theory lives in a probabilistic space. There may be something to learn, then, from studying how minimal substitution works when we lift General Relativity up into probabilistic space (as in the operational formulation in \cite{hardy2016operational}).

It is possible that we can find a quantum minimal substitution technique that mirrors very closely the classical minimal substitution technique once we lift up into probability space. In this process we will introduce into the quantum case something (call it the quantum metric) that plays an analogous role to the metric in the classical case. This will introduce additional degrees of freedom.

\subsection{Second element: addendum}

In QFT (before we apply the quantum minimal substitution technique) let there be $N_\text{QFT}$ real fields in $F_\mathbf{s}$ so we need this many equations to determine these fields from the specification of the operation (in terms of setting fields, outcome fields and their derivatives).   When we go to Quantum Gravity we have to introduce some quantum metric fields (analogous to the classical metric fields) and go over to a general quantum reference frame.  In so doing, we introduce extra fields. We need, therefore, extra equations (so now there will be a total of $N_\text{QG}>N_\text{QFT}$ real fields in $F_\mathbf{s}$). These will be the equations of Quantum Gravity that are analogous to the Einstein field equations.  My speculation is that, as we have lifted up into probability space, these will correspond to physicality conditions (that prevent future choices from influencing the past and keep probabilities bounded between 0 and 1).  To obtain such physicality conditions we can, by the quantum equivalence principle, transform into a quantum reference frame such that we have definite causal structure locally in the vicinity of some point. Then, at this point, we can impose physicality conditions as in QFT in the vicinity of this point.  Then we can transform back into the general quantum reference frame we are working in.

\subsection{Third element: interpretation}

The interpretation we propose (by strict analogy with General Relativity) is that the beables of Quantum Gravity are those quantities that are invariant under quantum diffeomorphisms. Since quantum reference frames are likely to be odd and counterintuitive objects, it is conceivable that the reality landscape will look quite different to the way it looks in standard Quantum Theory if we adopt this definition for beables. It is not inconceivable, then, that this will resolve the interpretational problems of Quantum Theory. Nonlocality (as witnessed by violation of Bell inequalities) may be nullified and the reality (or measurement) problem be solved in a natural way.


\part{Discussion}

\section{The game is afoot}

It took more than 200 years from Newton writing down his law of Universal Gravitation to Einstein adequately explaining where this came from.  We have lived with the puzzles of Quantum Theory for about 100 years now.  However, I do not believe it will take another 100 years before we make real progress.  Einstein had to solve the problem of Relativistic Gravity to come up with General Relativity.  It was in so doing that he removed the interpretational problems of Newton's theory of gravity. To formulate the problem of Relativistic Gravity it was necessary to have Special Relativistic Field Theories in place (such as electromagnetism).  There is a parallel between the problem of Relativistic Gravity and the problem of Quantum Gravity.  In the latter case, however, we have many of the pieces in place already.

It would be impossible to survey all work done on Quantum Gravity (and, certainly, I am not qualified to do this). The more established approaches are string theory \cite{polchinski1998string} and loop quantum gravity \cite{smolin2004invitation}.  Many in the string community have recently embraced quantum informational tools as a way of thinking about problems in quantum gravity (see for example \cite{ryu2006holographic, ryu2006aspects, susskind2016computational, swingle2012entanglement, faulkner2014gravitation}).  They have embraced the slogan \lq\lq it from qubit" - an adaptation of Wheeler's \lq\lq it from bit" due to Deutsch \cite{deutsch2004qubit} (see also \cite{d2015qubit}).  Hayden and Preskill's providing a quantum informational treatment of black holes \cite{hayden2007p} had a big impact on this way of thinking.   A number of the leading researchers in the loop community also pursue a strong interest in Quantum Foundations and this influences the work done in that field \cite{rovelli1996relational, smolin2012real, amelino2011principle}.  Other relatively established approaches to Quantum Gravity include the causal sets approach \cite{bombelli1987space} and the dynamical triangulations approach \cite{ambjorn2004emergence}.  The causal sets approach has strong quantum foundational inputs.  In particular, Sorkin and collaborators have developed the anhomomorphic logic \cite{sorkin2012logic} approach to address interpretational issues that arise (see also Sorkin's nice article \emph{Forks in the road, on the way to Quantum Gravity} \cite{sorkin1997forks}).

In the meantime, a broad approach to Quantum Gravity is beginning to take shape within a different section of the Quantum Foundations community based (among other places) in Oxford, Pavia, Vienna, Brussels, Morelia, Brisbane, and here in Waterloo.  Since this is such a broad approach (and, strictly speaking, not everyone is working directly on the problem of Quantum Gravity) it is difficult to give it a name. For the sake of convenience, let us call it \emph{Compositional Quantum Gravity} (though I am open to other names) since it has been born from attempting to understand Quantum Theory in compositional terms. This concerns reconstructions of Quantum Theory \cite{hardy2001quantum, dakic2009quantum, masanes2010derivation, chiribella2010probabilistic, chiribella2010informational, hardy2011reformulating, wilce2009four, wilce2016royal, goyal2008information, appleby2011properties, barnum2014higher, hoehn2014toolbox, hoehn2015quantum}, the General Probability Theories framework \cite{hardy2001quantum, barrett2007information}, the process framework \cite{abramsky2004categorical} (see the recent book \cite{coecke2017picturing}, the general boundaries framework \cite{oeckl2003general, oeckl2008general, oeckl2016local} (particularly its more recent incarnations \cite{oeckl2013positive, oeckl2014first, oeckl2016local}), studies of indefinite causal structure \cite{hardy2005probability, hardy2009quantum, chiribella2009beyond, chiribella2013quantum, oreshkov2012quantum, oreshkov2016operational, jia2018reduction, guerin2018observer, oreshkov2018whereabouts, baumeler2014perfect}, studies of correlations and inferential structure in Quantum Theory \cite{fritz2012beyond, wood2015lesson, henson2014theory, leifer2013towards}, quantum reference frames \cite{aharonov1967charge, aharonov1984quantum, bartlett2007reference, giacomini2017quantum}, and constructor theory \cite{deutsch2013constructor, deutsch2015constructor}.  This survey only scratches the surface - there is, by now, a considerable body of work along these lines.

One approach we may consider is to use an existing interpretation of Quantum Theory as a springboard to a theory of Quantum Gravity.  There is not much traffic in this direction.  Often the many worlds and \cite{saunders2010many} consistent (or decoherent) histories interpretations \cite{griffiths1984consistent, gell1993classical} are used to support ideas in Quantum Cosmology but this is not the same as using these interpretations to aid construction of a theory of Quantum Gravity.  There is some work connecting collapse models to Quantum Gravity in a phenological sense and this is a step in the right direction \cite{diosi1992quantum, penrose1996gravity}.  A few people have pursued issues in Quantum Gravity and Cosmology from within the de Broglie Bohm model (for example \cite{valentini2010inflationary, colin2016robust, de1997causal, pinto2005bohm}).

Our ultimate goal must be to understand the nature of reality. However we should, perhaps, temper our rush to get there.  If we find a natural and conceptually driven path to construct a theory of Quantum Gravity then I believe we are likely to arrive at such a deeper ontological understanding sooner.  Einstein's route to General Relativity is the supreme example of such an approach. Perhaps, guided by Einstein's spirit, we can achieve something similar.

\section*{Acknowledgements}

I am grateful to Wayne Myrvold for suggesting the Netwon-Cartan theory as a historical analogue in response to my exhortations that there is a danger in finding an interpretation of Quantum Theory when that theory will, likely, be replaced by a deeper theory of Quantum Gravity. That suggestion, unspun, resulted in the present paper.  I am grateful to Christopher Fuchs. In writing this paper I was strongly motivated by his inspirational \lq\lq Quantum Mechanics as Quantum Information (and only a little more)" \cite{fuchs2002quantum}.    I am grateful to Giacomo Mauro D'Ariano for inviting me to write a paper for a special issue. This was due a while back. I have probably missed the deadline.  I am grateful to Joy Christian for getting me interested in General Relativity and in Einstein's hole problem. It was also him who first told me about the Newton-Cartan formalism (which, incidentally, he quantized in \cite{christian1997exactly}). I am grateful to Antony Valentini for many discussions on reality and the occasional one on aether theories. I am grateful to: Doreen Fraser for discussions on theory construction; to Flaminia Giacomini and Esteban Castro-Ruiz for discussions on their quantum reference frame work (with Brukner); to Caslav Brukner for pointing out in response to an earlier draft that that his work with Gu\'erin provides a discrete realization of the quantum equivalence principle and, further, to Philippe Gu\'erin for discussions on this work; to Ding Jia for comments on an earlier draft, to Ognyan Oreshkov and Robert Oeckl for many discussions.  I am especially grateful to Zivy, Vivienne, and Helen for support and encouragement and to the people working at the Black Hole Bistro and at Aroma Cafe where most of this work was done.

Research at Perimeter Institute is supported by the Government of Canada through Industry Canada and by the Province of Ontario through the Ministry
of Economic Development and Innovation.  This project was made possible in part through the support of a grant from the John Templeton Foundation. The opinions expressed in this publication are those of the author and do not necessarily reflect the views of the John Templeton Foundation.  I am grateful also to FQXi for support through grant 2015-145578 (entitled "Categorical Compositional Physics") which funded some of this work.

\bibliography{QGbibJuly2018}
\bibliographystyle{plain}

\appendix

\part*{Appendix}

\addcontentsline{toc}{part}{Appendix}

\section{Physicality conditions in operator tensor QFT}\label{appendix}

\subsection{Basic Physicality Conditions}

Here we will consider physicality conditions for a general operation in operator tensor QFT.  We will write $\mathtt{a}^{\mathsf{T}_-}$ to indicate taking the input transpose so that the positivity condition in \eqref{QTphysicality} can be written as
\begin{equation}\label{physcondonenew}
0\leq \hat{A}_{\mathtt{a^{\mathsf{T}_-}} b^{\mathsf{T}_-}}^\mathtt{c^{\mathsf{T}_-}}
\end{equation}
We can write $\mathtt{a=a^+a^-}$ where $\mathtt{a}^+$ is the output part and $\mathtt{a}^-$ is the input part.  It is important to note that when $\mathtt{a}$ appears as a superscript then $\mathtt{a}^{\mathsf{T}_-}=\mathtt{a}_+(\mathtt{a}_-)^\mathsf{T}$ but when it appears as a subscript then, in that subscript position, we should write  $\mathtt{a}^{\mathsf{T}_-}=(\mathtt{a}_+)^\mathsf{T}\mathtt{a}_-$ (since, by notational convention, the inputs and outputs are reversed in the subscript position).  It is convenient to use the notation $u\leq^{\mathsf{T}_-}v$ meaning that the input transpose of $u$ is less than the input transpose of $v$.
Then we can write can write \eqref{physcondonenew} as
\begin{equation}
0\leq^{\mathsf{T}_-} \hat{A}_\mathtt{a b}^\mathtt{c}
\end{equation}
We write the other physicality condition in \eqref{QTphysicality} as
\begin{equation}
\hat{A}_\mathtt{a b}^\mathtt{c} \hat{I}^\mathtt{a_-b_-}_\mathtt{c_+} \leq^{\mathsf{T}_-} \hat{I}_\mathtt{a_+ b_+}^\mathtt{c_-}
\end{equation}
Note that the operators $\hat{I}^\mathtt{a_-b_-}_\mathtt{c_+}$ and $\hat{I}_\mathtt{a_+ b_+}^\mathtt{c_-}$ have only input wires (because, by convention, inputs and outputs are reversed in subscripts). Deterministic operations (discussed below) saturate this inequality.

\subsection{Issues}

This is not, however, a complete set of physicality conditions because output wires of $\mathsf{A}_\mathtt{a b}^\mathtt{c}$ can be wired forward so that they feed into input wires of this same operation.  To get a complete set of physicality conditions we need to adapt the techniques of Chiribella, D'Ariano, and Perinotti (CDP)\cite{chiribella2009theoretical} in their paper on quantum combs. There they provide a recursive method to place constraints on \lq\lq combs" (combs have this same property that outputs can be fed into inputs on the same comb).  There is a dictionary between the operator tensor approach to Quantum Theory (as discussed in \cite{hardy2011reformulating}) and the quantum combs approach explained in \cite{hardy2011reformulating}.  QFT introduces the additional problem that do not have discreteness (combs have both a finite number of teeth and the Hilbert spaces are finite).  This finiteness is important for the quantum combs approach because (i) it guarantees that the recursion terminates and (ii) the identity is well defined in such a case. We will propose a limiting technique to get round these problems for operator tensor QFT.

\subsection{Deterministic operations}

\subsubsection{Preliminaries}

A deterministic operation is one whose set of outcomes is the set of all possible outcomes. Any circuit comprised of only deterministic operations must have probability one.  We will first show how to test whether deterministic operators are physical. We can then give conditions for general operators (though we leave it as an open question as to whether this is a complete set of conditions). 

A deterministic operation having only inputs must correspond to the identity operator because of causality. For example, for any $\mathtt{F}^\mathtt{a_-}[\text{det}]$, we have
\begin{equation}\label{paviacausality}
\hat{F}^\mathtt{a_-}[\text{det}] = \hat{I}^\mathtt{a_-}
\end{equation}
If we could have different operators, $\hat{F}^\mathtt{a_-}[\text{det}]$ and $\hat{G}^\mathtt{a_-}[\text{det}]$, corresponding to different deterministic operations, $\mathsf{F}^\mathtt{a_-}[\text{det}]$ and $\mathsf{G}^\mathtt{a_-}[\text{det}]$, then we could send information back in time since we would have $\hat{A}_\mathtt{a_-} \hat{F}^\mathtt{a_-}[\text{det}] \not= \hat{A}_\mathtt{a_-} \hat{G}^\mathtt{a_-}[\text{det}]$. This would violate causality because then the probability of the outcome associated with $\mathtt{A}_\mathtt{a-}$ (in the past) would depend on what we did in the future.  This causality argument demonstrates that deterministic effects must be equal (this is the Pavia causality condition \cite{chiribella2010informational}).  To see that they must be equal to the identity it is sufficient to look at a measurement whose outcomes correspond projectors onto an eigenbasis.  If we ignore the outcome we get the sum of the corresponding projectors which is equal to the identity.

Consider a deterministic operation, $\mathtt{A}_\mathtt{c}^\mathtt{d}[\text{det}]$, that can be written
\begin{equation}\label{EdetCdetDdet}
\mathtt{A}_\mathtt{c}^\mathtt{d}[\text{det}]=\mathtt{C}^\mathtt{h^+}_\mathtt{c}[\text{det}] \mathtt{D}_\mathtt{h^+}^\mathtt{d}[\text{det}]=
\begin{Compose}{0}{-1.2}
\ccircle{C}{2.1}{0,0}\csymbol{C[\text{det}]}\ccircle{D}{2.1}{1,7} \csymbol{D[\text{det}]}
\joincc[right]{C}{80}{D}{-100} \csymbol[10,0]{h^+}
\thispoint{Cin}{5,-2} \joincc[above]{Cin}{180}{C}{-10} \csymbol{c}
\thispoint{Dout}{-3,8} \joincc[above]{D}{170}{Dout}{-10} \csymbol{d}
\end{Compose}
\end{equation}
The important point is that it can be partitioned into a past and a future part by a spacelike typing surface.  A (forward pointing) spacelike typing surface is one for which $\mathtt{h}=\mathtt{h^+}$ and we denote it by $\mathtt{h^+}$.  The typing surfaces, $\mathtt{c}$ and $\mathtt{d}$ may have both input and output parts.   Causality demands that
\begin{equation}\label{detcondition}
\hat{A}_\mathtt{c}^\mathtt{d}[\text{det}] \hat{I}_\mathtt{d+} = \hat{I}^\mathtt{d-} \hat{B}_\mathtt{c}[\text{det}]
\end{equation}
where $\hat{B}_\mathtt{c}[\text{det}]$ is, itself, an allowed operator corresponding to a deterministic operation satisfying
\begin{equation}\label{Bdet}
\hat{B}_\mathtt{c}[\text{det}] = \hat{C}^\mathtt{h_+}_\mathtt{c}[\text{det}]\hat{I}_\mathtt{h_+}
\end{equation}
(the 1 is to denote the first step in a recursive sequence as discussed below).
This is because, by \eqref{paviacausality}
\begin{equation}
\hat{I}_\mathtt{d+} \hat{D}_\mathtt{h_+}^\mathtt{d}[\text{det}] = \hat{I}_\mathtt{h_+}^\mathtt{d_-} = \hat{I}_\mathtt{h_+} \hat{I}^\mathtt{d_-}
\end{equation}
Note that $\hat{I}^\mathtt{d_-}$ and $\hat{I}_\mathtt{h_+}$ have input wires (they are deterministic effects).
Equation \eqref{detcondition} gives us a constraint on operators associated with deterministic operations.  Diagrammatically, we can describe the condition in \eqref{detcondition} as
\begin{equation}\label{detconditiondiagram}
\begin{Compose}{0}{-1.2} \setdefaultfont{\mathnormal}
\ccircle{C}{2.1}{0,0}\csymbol{\hat{C}[\text{det}]}\ccircle{D}{2.1}{1,7} \csymbol{\hat{D}[\text{det}]}
\joincc[left]{C}{80}{D}{-100} \csymbol{h_+}
\thispoint{Cin}{5,-2} \joincc[above]{Cin}{180}{C}{-10} \csymbol{c}
\thispoint{Doutm}{-3,8} \joincc[below left]{D}{170}{Doutm}{-10} \csymbol{d_-}
\ucircle{dI}{-3,11} \csymbol{\hat{I}}\joincc[above right]{D}{135}{dI}{-45} \csymbol{d_+}
\end{Compose}
~~=~~
\begin{Compose}{0}{-0.8}\setdefaultfont{\mathnormal}
\ccircle{C}{2.1}{0,0}\csymbol{\hat{C}[\text{det}]}  \ucircle{hI}{1,5} \csymbol{\hat{I}} \joincc[left]{C}{80}{hI}{-110} \csymbol{h_+}
\thispoint{Cin}{5,-2} \joincc[above]{Cin}{180}{C}{-10} \csymbol{c}
\ucircle{dI}{-1, 6} \csymbol{\hat{I}} \thispoint{dIout}{-4,8} \joincc[above]{dI}{135}{dIout}{-45} \csymbol{d_-}
\end{Compose}
\end{equation}
where we have substituted in \eqref{EdetCdetDdet} on the left and \eqref{Bdet} on the right to make it clearer where this condition comes from.

This condition for the deterministic case is used to construct conditions for the general case.

\subsubsection{Recursion}

The condition in \eqref{detcondition} (shown diagrammatically in \eqref{detconditiondiagram}) must be applied recursively since, to check it, we need to check that $\hat{B}_\mathtt{c}[\text{det}]$ is physical.  To do this we need to introduce a spacelike hypersurface across region $\mathtt{B}$ and so on.  To notate this let us change our notation so
\[ \mathtt{A} = \mathtt{A}[0], ~~~~\mathtt{B} = \mathtt{A}[1],~~~~ \mathtt{h}^+ = \mathtt{h}^+[1], ~~~~\mathtt{c}=\mathtt{c}[1],~~~~ \mathtt{d}=\mathtt{d}[1] \]
At each subsequent step in the recursion we increment the integer in square brackets by 1.  In a general step, $n$, we need to check the conditions
\begin{equation}
0\leq^{T-} \hat{A}_{\mathtt{c}[n]}^{\mathtt{d}[n]}[n-1,\text{det}], ~~~~~~
\hat{A}_{\mathtt{c}[n]}^{\mathtt{d}[n]}[n-1,\text{det}] \hat{I}_{\mathtt{d^+}[n]} = \hat{I}^{\mathtt{d^-}[n]} \hat{A}_{\mathtt{c}[n]}[n,\text{det}]
\end{equation}
where
\[  \mathtt{c}[n]=\mathtt{c}[n+1]\mathtt{d}^R[n+1] \]
as we iterate to the next step  (with $\mathtt{c}[n+1]$ to the past of $\mathtt{h}^+[n+1]$ and $\mathtt{d}[n+1]$ to the future).  This sequence will terminates on the $N$th step if
$\mathtt{c}[N] = \mathtt{c}^-[N]$ so that then region $\mathtt{A}[N]$ has only an input. For this step we check that
\begin{equation}
0\leq^{T-} \hat{A}_{\mathtt{c}[N]}^{\mathtt{d}[N]}[N-1,\text{det}], ~~~~~~
\hat{A}_{\mathtt{c}[N]}[N,\text{det}] = \hat{I}_{\mathtt{c}[N]}[N,\text{det}]
\end{equation}
This adapts the techniques of CDP the present situation.  However, as we will now see, it is not so simple here.

\subsubsection{Challenges}

Quantum combs have finite number of teeth (corresponding to a particular  finite foliation) and, further, the Hilbert spaces are finite dimensional so the recursive method is guaranteed to terminate and also there are no tricky technical problems in defining the identity operator.  In the present case we have continuous dimensional Hilbert spaces and this introduces a number of challenges: (i) first there is not a unique choice for this sequence of \lq\lq hypersurfaces"; (ii) we can have an arbitrarily large number of such hypersurfaces; (iii) for general regions, $\mathtt{A}$, the recursion will not terminate as there will always be some output component; (iiii) since we have continuous dimensional Hilbert spaces we do have some tricky issues defining the identity operator.

\subsubsection{Sketch of way forward}

To make progress we proceed in the following way.
\begin{description}
\item[Foliation generation.]  We introduce a number of points across and outside the region $\mathtt{A}$. This could be done by sprinkling \cite{sorkin2005causal} or by some regular grid. We let this grid or sprinkling have some characteristic length $l$. Then we form the causal set \cite{sorkin2005causal} (by referring to the metric) from these points. Next we find foliations of region $\mathtt{A}[n]$ whose hyperplanes pass through these wires of the causal set (using the causal set structure to guide this construction).
\item[Finite approximation.] We approximate the operator, $\hat{A}_{\mathtt{c}[n]}[n,\text{det}]$, with some operator, $\hat{A}_{\mathtt{c}[n,K_n]}[n,\text{det}, K_n]$, corresponding to a finite Hilbert space as described below.  We define a measure, $A_l$, of how good this approximation is (with $A_l=0$ when the approximation is perfect).  We demand of this approximation, that $A_l \rightarrow 0$ as $l \rightarrow 0$.
\item[Physicality witness.] We define a physicality witness, $W_l$, which measures how well the physicality conditions are satisfied (with $W_l=0$ when they are perfectly satisfied).  We say that $\hat{A}_\mathtt{c}^\mathtt{d}[\text{det}]$ is physical if $W_l \rightarrow 0$ as $l \rightarrow 0$.
\end{description}
To make this a little more formal, put $l=L/m$ for $m=1, 2, 3, \dots$ and now define $A^m=A_{L/m}$ and $W^m=W_{L/m}$.  We demand that the sequence
\[ A^1, A^2, A^3, \dots \]
converges to zero.  Then we say that $\hat{A}_\mathtt{c}^\mathtt{d}[\text{det}]$ is physical if the series
\[ W^1, W^2, W^3, \dots \]
converges to zero.

Certainly the set of conditions we have considered is a necessary set for physicality. We haven't strictly shown that it is a sufficient set.  Further, we have not proven the converse - namely that if an operator satisfies this set of conditions then it can built out of physical operators corresponding to the microscopic regions.  We will leave these matters for future work and simply conjecture that this is a sufficient set of conditions for physicality.

It also may be that the set of conditions we have considered is overcomplete.  Foliations have relationships between them and it may not be necessary to include all foliations as we are doing.

There may be many ways of implementing the basic strategy outlined above.  I will now discuss in some detail one way we might do this.

\subsubsection{Foliation generation}

We wish to generate sets of foliation.  First we choose some given characteristic length $l$. Then we sprinkle or set up a grid of points according to $l$.  Next we form the causal set.  Now we obtain a foliation for the region $\mathtt{A}$ employing this causal set by means of spacelike hyperplanes passing through the wires of the causal set (such that every wire is included in at least one hyperplane and only one hyperplane for any given set of wire intersections is included in the foliation).  Any such set of hyperplanes will be finite.
We keep the last $N$ of this set where the first hyperplane we keep is the last one having the property that it partitions region $\mathtt{A}$ into two regions such that the earlier of these two regions has only inputs. We do this because the recursion process (which is applied backwards in time) will terminate at the step involving this hypersurface. There will be a rough relationship between $N$ and $l$ of the form
\[ N \propto \frac{1}{l}  \]
i.e. as we decrease $l$ we will have more hyperplanes in our foliation.

For any given $l$ (and consequent given causal set) we can generate many foliations.  We only keep one representative example for each particular pattern of intersections of hypersurfaces and the wires of the causal sets. Since, then, the causal set has a finite number of such wires, the total number of foliations we generate will be finite also.

\subsubsection{Approximation by finite dimensional Hilbert spaces}

We assume we can regard the continuous dimensional Hilbert spaces, $\mathcal{H}_{\mathtt{c}[n]}$ and $\mathcal{H}_{\mathtt{d}[n]}$, encountered in step $n$ as each being spanned by a basis of orthonormal eigenstates $\{|\varphi^i[n,l]\rangle_{\mathtt{c}[n]}: i=0~\text{to}~ \infty\}$ and $\{|\varphi^j[n,l]\rangle_{\mathtt{d}[n]}: j=0~\text{to}~ \infty\}$.  For step $n$ we choose some finite number, 
\[K_n=D2^{(N-n+1)}\]
and project the operators down to the first $K_n$ states in these orthonormal bases (here $D$ is a constant we choose corresponding to the number of qubits per foliation increment).  In this way we obtain operators
\[ \hat{A}_{\mathtt{c}[n,K_n]}^{\mathtt{d}[n,K_n]}[n-1,\text{det},K_n ], ~~~ \hat{I}_{\mathtt{d^+}[n,K_n]} ~~~ \hat{I}^{\mathtt{d^-}[n,K_n]} ~~~\hat{A}_{\mathtt{c}[n,K_n]}[n,\text{det},K_n]   \]
We define the quantity
\begin{equation}
\alpha_{nlf}= ||\hat{A}_{\mathtt{c}[n]}[n,\text{det}] -\hat{A}_{\mathtt{c}[n]}[n,\text{det},K_n]   ||
\end{equation}
where $f$ labels the particular foliation for the causal set associated with this $l$ and
where $||\cdot ||$ is some appropriate norm (such as the trace norm).  For each causal set (for some given  $l$ ) we define
\begin{equation}
A_l = \frac{1}{F}\sum_{nf}  \frac{\alpha_{nlf} }{K_n}
\end{equation}
where $F$ is the number of foliations for this causal set. This is a measure of how badly the operators are approximated for any given causal set.

We demand that $A_{l}$ tends to zero as $l\rightarrow 0$. This constraint can be thought of as a constraint on the class of operators we consider.  It is also a constraint on how we choose our set of orthonormal eigenstates. In particular, we expect that they correspond to a set of orthonormal functions whose frequency increases with $i,j$.

\subsubsection{Physicality witness.}

For the causal set associated with $l$ we check for physicality by checking the following physicality conditions (these are on the truncated Hilbert spaces) for each $n$ for each foliation
\begin{equation}
0\leq^{T-} \hat{A}_{\mathtt{c}[n,K_n]}^{\mathtt{d}[n,K_n]}[n-1,\text{det},K_n]
\end{equation}
\begin{equation}\label{causalitycondfinite}
\hat{A}_{\mathtt{c}[n,K_n]}^{\mathtt{d}[n,K_n]}[n-1,\text{det},K_n] \hat{I}_{\mathtt{d}^+[n,K_n]} = \hat{I}^{\mathtt{d^-}[n,K_n]} \hat{A}_{\mathtt{c}[n]}[n,\text{det},K_n]
\end{equation}
(though for the last step, we check 
\begin{equation}\label{finalstepfinite}
0\leq^{T-} \hat{A}_{\mathtt{c}[N,K_n]}^{\mathtt{d}[N,K_n]}[N-1,\text{det},K_n], ~~~~~~
\hat{A}_{\mathtt{c}[N,K_n]}[N,\text{det},K_n] = \hat{I}_{\mathtt{c}[N,K_n]}[N,\text{det},K_n]
\end{equation}
instead). 
In general we expect these conditions to fail. We can take some measure of how much they fail and take the average of these to get a measure for how much physicality fails for this given $l$.  A measure of how much the first condition fails is given by taking the sum of the negative eigenvalues. Call this $\text{neg}_{nlf}$ (where $f$ denotes the particular foliation).  To get a measure of how much the second condition fails we can write
\begin{equation}\label{detconditionfailure}
\hat{A}_\mathtt{c}^\mathtt{d}[n, \text{det}, K_n] \hat{I}_{\mathtt{d+}[n,K_n]} = \hat{I}^{\mathtt{d-}[n,K_n]} \hat{A}_{\mathtt{c}[n,K_n]}[n, \text{det}, K_n] + \hat{A}_{\mathtt{c}[n,K_n]}^{\mathtt{d}[n,K_n]}[n,\text{fail}, K_n]
\end{equation}
where $\hat{A}_{\mathtt{c}[n,K_n]}^{\mathtt{d}[n,K_n]}[n,\text{fail}, K_n]$ contains no support on $\hat{I}^{\mathtt{d-}[n,K_n]}$ (we can write a similar expression for the final step in \eqref{finalstepfinite}).  Then the failure of physicality, at this step is given by
\[ \beta_{nlf}= ||\hat{A}_{\mathtt{c}[n,K_n]}^{\mathtt{d}[n,K_n]}[n,\text{fail}, K_n] || \]
We take the average of each of these failures to get a physicality witness
\begin{equation}
W_l = \frac{1}{F}\sum_{nf}  \frac{|\text{neg}_{nlf}|+\beta_{nlf}}{K_n}
\end{equation}
Clearly, if all the physicality conditions were satisfied this would equal zero.

\subsection{General Operators}

Once we can say whether an operator is physical for the deterministic case then we can look at the general case (where the associated operations correspond to a subset of all possible outcomes).  Necessary conditions for the operator $\hat{A}_\mathtt{c}^\mathtt{d}$ to be physical are that (i) it has positive input transpose and (ii) that there exists another operator (associated with the same region) with positive input transpose such that the sum of these two operators is equal to a deterministic operator for this region.  In the analogous situation concerning quantum combs CDP proved that this is also sufficient.  We will not attempt a prove of sufficiency here but rather conjecture that this is also a sufficient condition.  There is at least one cloud on the horizon however. In particular, for a very wide region with only inputs there is a limit to how entangled this effect can be. It is not clear how this is captured in the conditions given here.

\subsection{General issues}

The ideas outlined here are rather tentative and some conjecture is involved at this stage.  Certainly we expect there to exist a good set of physicality conditions. It would be good if these were, further, easy to check. In the particular approach above we employ a cutoff and it would be interesting to see how the usual issues of UV divergence in QFT look in this approach.  There may be better techniques to test whether the condition \eqref{detcondition} is always satisfied employing symmetry and rescaling arguments for example.

\end{document}